\documentclass[onecolumn,aps]{revtex4}
\usepackage{amssymb}
\newcommand{\sect}[1]{\setcounter{equation}{0}\section{#1}}

\newcommand{\bfm}[1]{\mbox{\boldmath${#1}$}}

\begin{document}

\title{\bf Gauge equivalence among quantum\\ nonlinear many body systems}

\author{Antonio M. \surname{Scarfone}}

\affiliation{Istituto
Nazionale di Fisica della Materia (CNR-INFM) and\\ Dipartimento di Fisica --
Unit\'a del Politecnico di Torino,\\ Corso Duca degli Abruzzi 24,
I-10129 Torino, Italy.}
\date{\today}

\begin {abstract}
Transformations performing on the dependent and/or the independent variables
are an useful method used to classify PDE in class of equivalence.
In this paper we consider a large class of U(1)-invariant nonlinear Schr\"odinger
equations containing complex nonlinearities. The U(1) symmetry
implies the existence of a continuity equation for the particle
density $\rho\equiv|\psi|^2$ where the current ${\bfm j}_{_\psi}$
has, in general, a nonlinear structure. We introduce a nonlinear
gauge transformation on the dependent variables $\rho$ and ${\bfm j}_{\psi}$
which changes the evolution equation in another
one containing only a real nonlinearity and transforms the particle
current ${\bfm j}_{_\psi}$ in the standard bilinear form. We extend
the method to U(1)-invariant coupled nonlinear Schr\"odinger
equations where the most general nonlinearity is taken into account
through the sum of an Hermitian matrix and an anti-Hermitian matrix.
By means of the nonlinear gauge transformation we change the
nonlinear system in another one containing only a purely Hermitian
nonlinearity. Finally, we consider nonlinear Schr\"odinger equations
minimally coupled with an Abelian gauge field whose dynamics is
governed, in the most general fashion, through the
Maxwell-Chern-Simons equation. It is shown that the nonlinear
transformation we are introducing can be applied, in this case,
separately to the gauge field or to the matter field with the same
final result. In conclusion, some relevant examples are presented to
show the applicability of the method.\\

\end {abstract}
\maketitle

\sect{Introduction} The Schr\"odinger equation is one of the most
studied topics both from a mathematical and physical point of
wiev. A particular interest is related to the possible nonlinear
extensions of this equation. Just one year after the discovery of
the Schr\"odinger equation, Fermi proposed
its first nonlinear generalization \cite{Fermi}.\\
In the following years, many nonlinear extensions of this equation
have been proposed in literature in order either to explore
fundamental arguments of the quantum mechanics, with the usual
linear theory representing only an approximation, or to describe
particular phenomenological physical effects. Among the many
attempts made to generalize in a nonlinear manner the Schr\"odinger
equation we recall the Bialynicki-Birula and Mycielski equation
\cite{Bialynicki}, with the nonlinear term
$-b\,\ln(|\psi|^2)\,\psi$; the Guerra and Pusterla model
\cite{Guerra}, that with the purpose of preserving the superposition
principle of the quantum mechanics introduced the nonlinear term
$(\Delta |\psi|/|\psi|)\,\psi$; more recently, the Weinberg model
\cite{Weinberg,Weinberg1}, with the introduction of homogeneous
nonlinear terms in order to save partially the same fundamental
principle.\\ On a phenomenological basis we recall the well known
cubic Schr\"odinger equation \cite{Gross2,Gross1,Pitaevskii}, used
in the study of the dynamical evolution of a Boson gas with a
$\delta$-function pair-wise repulsion or attraction
\cite{Barashenkov} and in the description of the Bose-Einstein
condensation of alcali atoms like $^7$Li, $^{23}$Na and $^{87}$Rb
\cite{Shi,Stringari}; the model introduced by Kostin
\cite{Kostin2,Kostin1,Schuch2,Schuch1}, with the nonlinear term
$i\,\ln(\psi/\psi^\ast)\psi+i\,\langle\ln(\psi/\psi^\ast)\rangle\,\psi$
used to describe dissipative systems, and others
\cite{Dodonov,Gisin,Grigorenko,Malomed,Martina}.\\
Many nonlinear Schr\"odinger equations (NLSEs) contain complex
nonlinearities. For instance, a nonlinearity of the type
$a_1\,|\psi|^2\,\psi+a_2\,|\psi|^4\,\psi+
i\,a_3\,\partial_x(|\psi|^2\,\psi)+(a_4+i\,a_5)\,\partial_x|\psi|^2\,\psi$
was introduced to describe a single mode wave propagating in a Kerr
dielectric guide \cite{Gagnon2,Gagnon1}; the nonlinearity
$a_1\,|\psi|^2\,\psi+i\,a_2\,\psi+
i\,a_3\,\partial_{xx}\psi+i\,a_4\,|\psi|^2\,\psi$, proposed in
\cite{Malomed2,Malomed1} to take into account of pumping and dumping
effects of the nonlinear media, is used to describe dynamical modes
in plasma physics, hydrodynamics, and also solitons in optical
fibers (\cite{Malomed3} and references therein); the nonlinearity
$a_1\,|\psi|^2\,\psi+
i\,a_2\,\partial_{xxx}\psi+i\,a_3\,\partial_x(|\psi|^2\,\psi)
+i\,a_4\,\partial_x|\psi|^2\,\psi$ introduced to describe
the propagation of high power optical pulses in ultra-short soliton
communication systems
\cite{Gedalin,Karpman,Karpman1,Karpman2,Karpman3,Li,Doktorov}. In
\cite{Kaniadakis,Kaniadakis1} a NLSE with the complex nonlinearity
$\kappa\,(\psi^\ast\,{\bfm\nabla}\psi-\psi\,{\bfm\nabla}\psi^\ast)\,
{\bfm\nabla}\,\psi+(\kappa/2)\,{\bfm\nabla}(\psi^\ast\,{\bfm\nabla}
\psi-\psi\,{\bfm\nabla}\psi^\ast)\,\psi$ has been introduced to take
into account a generalized Pauli exclusion-inclusion principle
between the quantum particles constituting the system. In
\cite{Scarfone7}, in the stochastic quantization framework, starting
from the most general kinetic containing a nonlinear drift term and
compatible with a linear diffusion term, a class of NLSEs with a
complex nonlinearity was derived whilst, recently
\cite{Scarfone10,Scarfone8,Scarfone9}, a wide class of NLSEs has been obtained
starting from the quantization of a classical many body system whose
underlying kinetic is described by a nonlinear Fokker-Planck
equation associated to a generalized trace-like entropy. Finally,
the Doebner-Goldin equation
\cite{Doebner3,Doebner2,Doebner1,Goldin1} was introduced from
topological considerations as the most general class of NLSEs
compatible with the linear Fokker-Planck equation for the
probability density $\rho\equiv|\psi|^2$, where the nonlinear term
was derived from the unitary group representation of the
infinite-dimensional diffeomorphism group proposed as a
{\sl universal quantum kinematical group} \cite{Goldin2}.\\
In the recent years, an increasing interest has been also addressed
to systems of coupled nonlinear Schr\"odinger equations (CNLSEs),
particularly after the invention of high-intensity lasers which have
allowed \cite{Mollenauer} the experimental test of the pioneering
theoretical works on the optical fibers propagation in long-distance
communications \cite{Hasegawa,Tappert}. In fact, single-mode optical
fibers are not really single-mode type since two possible
polarizations exist. A rigorous study of their propagation requires
the use of CNLSEs in order to take into account the evolution of the
different polarized waves. In 1974 Manakov \cite{Manakov} introduced
a CNLSE starting from the cubic NLSE by considering that the total
field is a superposition of two, left and right polarized fields.
When ultrashort pulses are transmitted through fibers, CNLSEs with
complex high derivative nonlinearities arise
\cite{Mahalingam,Nakkeeran1,Nakkeeran2,Nakkeeran3,Lakshmanan,Vinoj}.
CNLSEs are also employed in the study of light propagation through a
nonlinear birefringent medium, in systems with nonrelativistic interactions
among the different kind of particles, in spinor Bose-Einstein
condensation, in the description of micro-polar elastic solids,
among the many
\cite{Ablowitz,Agrawal,Zakharov,Erbay,Bose,Matthews,Newboult,Ryskin,Yip}.\\
Finally, when coupled with gauge fields, NLSEs are useful in the
study of some interesting phenomenologies in condensed matter
physics. For instance, in the Ginzburg-Landau theory of the
superconductivity the cubic NLSE is coupled with an Abelian gauge
field whose interaction is described by means of the Maxwell
equations \cite{Kaper,Tomaras}. Some time, gauge field dynamics can
be described by the further Chern-Simons term which confers mass to
the field without destroy the gauge invariance of the theory. These
models have particle-like solutions obeying to a non-conventional
statistics, named anyons \cite{Jackiw3,Jackiw1,Wilczek1},
which can find an application in the study of high-$T_{\rm c}$
super-conductors \cite{Wainberg2,Wilczek2}.\\
In this work we present, in a unify way, some recent results
concerning the gauge equivalence among U(1)-invariant
nonrelativistic quantum systems described by nonlinear Schr\"odinger
equations containing a complex nonlinearity
\cite{Scarfone4,Scarfone3,Scarfone2,Scarfone1,Scarfone5}. We
discuss, in particular, some aspects of the gauge transformations of
the third kind based on its capability to change the NLSE in another
one containing only a real nonlinearity (in the case of CNLSE
complex nonlinearities are replaced by non Hermitian nonlinearities
and the gauge transformation changes them in purely Hermitian
nonlinearities). It is worthy to observe that nonlinear
transformations of the type discussed in the following have been
introduced previously in literature to relate different families of
NLSEs \cite{Kundu2,Kundu1}, CNLSEs \cite{Wadati,Wadati2} and, more in general,
nonlinear PDE \cite{Calogero6,Calogero5,Calogero4a,Calogero4}.\\
In \cite{Doebner4} the name of ``gauge transformations of third
kind'' has been coined for the class of the unitary nonlinear
transformations. They differ from the gauge transformations of first
kind, which have constant generators, and those of the second kind,
which have generators depending on the space coordinate and
eventually on time, since, the gauge transformations of third kind
have generators depending functionally, often in
a nonlinear manner, on the fields.\\
On the physical ground, they are named {\em gauge transformations}
because, as stated by Feynmann and Hibbs \cite{Feynmann}, in a
nonrelativistic quantum mechanics, all measurements of observables
are always accomplished through a measurement of position and time.
Thus, quantum theories, for which the corresponding wave-functions
give the same probability density in space at all time are in
principle equivalent \cite{Doebner4}. In particular, when the
wave-functions $\psi$ and $\phi$ are related each to the other by a
unitary transformation, like in the gauge transformations which we
are introducing, the densities of probability of position coincide
$|\psi|^2=|\phi|^2\equiv\rho$ and the fields
$\psi$ and $\phi$ describe the same system. In this sense,
nonlinear gauge transformations permit to classify NLSEs in classes
of equivalence. Any member belonging to the same class, in spite of
its nonlinearity, describe the same physical system.\\
After the brief introductory next section about the notations used in this
paper we begin, in section 3, by considering a wide class of
canonical NLSEs with complex nonlinearity in a $n+1$-dimensional
space-time (throughout this paper we use unities $\hbar=c=e=1$ and
we set $m=1/2$)
\begin{equation}
i\,\frac{\partial\psi}{\partial
t}+\Delta\psi+\Big(W[\psi^\ast,\,\psi]+i\,{\mathcal
W}[\psi^\ast,\,\psi]\Big)\,\psi=0 \ ,\label{sch1}
\end{equation}
describing, in the mean field approximation, the dynamics of a
nonrelativistic scalar field $\psi$ conserving the particle number
$N=\int\rho\,d^nx$. The real $W[\psi^\ast,\,\psi]$ and the imaginary
${\mathcal W}[\psi^\ast,\,\psi]$ nonlinearities appearing in
equation (\ref{sch1}) are smooth functionals of the fields
$\psi,\,\psi^\ast$ and of their spatial derivatives of any order.
Since the nonlinearity in the evolution equation is complex the
particle current ${\bfm j}_{_\psi}$ has, in general, a nonlinear
structure which differs from the bilinear form of the standard
linear quantum mechanics. We introduce the Lagrangian formulation
both in the wave-function representation and in the hydrodynamic
representation and we study the U(1) symmetry, which plays a
relevant role for the purpose of the present work. Than, we
introduce a nonlinear unitary transformation $\psi\to\phi$ that
changes the complex nonlinearity $W[\psi,\,\psi^\ast]+i\,{\mathcal
W}[\psi,\,\psi^\ast]$ in another one $\widetilde
W[\phi,\,\phi^\ast]$ which turns out to be purely real. As a
consequence the new current ${\bfm j}_{_\phi}$ assumes the bilinear
form of the linear Schr\"odinger theory. In
\cite{Calogero3,Calogero6,Calogero5,Calogero4a,Calogero4,Calogero1,
Doebner4,Fordy,Hisakado2,Hisakado1,Jackiw,Kundu2,Kundu1,Wadati,Wadati2} we can find
some examples of nonlinear gauge transformation of the third kind.
All transformations introduced there can be systematically obtained
with the method here presented.\\
In section 4, we generalize the transformation to CNLSEs by
considering the following system of equations
\begin{equation}
i\,\frac{\partial\Psi}{\partial t}+\hat{A}\,
\Delta\Psi+\Big(\widehat W[\Psi^\dag,\,\Psi]+i\,\widehat{\mathcal
W}[\Psi^\dag,\,\Psi]\Big)\,\Psi=0 \ ,\label{csch}
\end{equation}
where $\Psi=(\psi_{_1},\,\ldots,\,\psi_{_p})$ is a $p$-dimensional
vector of scalar wave-functions and the nonlinearity
$\widehat\Lambda[\Psi^\dag,\,\Psi]=\widehat
W[\Psi^\dag,\,\Psi]+i\,\widehat{\mathcal W}[\Psi^\dag,\,\Psi]$ is
composed by an Hermitian matrix $\widehat
W=(\widehat\Lambda+\widehat\Lambda^\dag)/2$ and an anti-Hermitian
matrix $i\,\widehat{\mathcal
W}=(\widehat\Lambda-\widehat\Lambda^\dag)/2$. We assume that the
system (\ref{csch}) has $q\leq p$ conserved multiplets which implies
the existence of $q$ continuity equations. Then, we introduce a
nonlinear gauge transformation $\Psi\rightarrow\Phi$ which
transforms the CNLSE in another one with a purely Hermitian
nonlinearity $\widehat\Lambda^\prime[\Psi^\dag,\,\Psi]=\widehat
W^\prime[\Psi^\dag,\,\Psi]$. As a consequence, the transformed
currents assume the standard bilinear form. \\
Finally, in section 5, we generalize the method to a class of NLSEs
minimally coupled with an Abelian gauge field $A_\mu$, where the
matter field is described by the following equation
\begin{equation}
i\,D_0\psi+{\bfm D}^2\psi+\Big(W[\psi^\ast,\,\psi,\,\bfm{
A}]+i\,{\mathcal W}[\psi^\ast,\,\psi,\,\bfm{ A}]\Big)\,\psi=0 \
,\label{schroedinger}
\end{equation}
with $\psi$ the scalar charged field and
$D_\mu\equiv(D_0,\,{\bfm D})$ denotes the standard covariant
derivative. The dynamics of the gauge field is provided by the
Maxwell-Chern-Simons equation
\begin{equation}
\gamma\,\partial_\mu F^{\mu\nu}+g\,\varepsilon^{\nu\tau\mu}
\,F_{\tau\mu}=j^\nu_{_{A\psi}} \ ,\label{ggg}
\end{equation}
where $F_{\mu\nu}=\partial_\mu A_\nu-\partial_\nu A_\mu$ is the
electromagnetic field and $j^\nu_{_{A\psi}}\equiv(\rho,\,{\bfm
j}_{_{A\psi}})$ is the covariant current. We stress one more that
the charged current ${\bfm j}_{_{A\psi}}$ has, in general, a
nonlinear structure due to the presence, in the evolution equation
(\ref{schroedinger}), of the complex nonlinearity
$W[\psi^\ast,\,\psi,\,\bfm{ A}]+i\,W[\psi^\ast,\,\psi,\,\bfm{ A}]$.
The gauge transformation can be applied equivalently to the matter
field or to the gauge field obtaining the same final result: the
nonlinearity in the Schr\"odinger equation (\ref{schroedinger})
turns out to be purely real and the expression of the charged
current reduces to
the standard bilinear form.\\
In section 6, we collect some explicit examples to illustrate the
applicability of the method whilst, in the conclusive section 7, we
discuss the possible further development about the nonlinear gauge
transformations presented in this paper.
\section{Preliminary mathematical background}

Let $M$ be a complex $n$-dimensional smooth manifold labeled by the
vector ${\bfm x}\equiv(x_{_1},\,\ldots,\,x_{_n})$. Let ${\mathcal
F}:\,\,M\rightarrow I\!\!R$ be the algebra of the functions on $M$
and $F:\,\,{\mathcal F}\rightarrow I\!\!R$ the algebra of the
functionals on ${\mathcal F}$ of the type $G=\int{\mathcal G}({\bfm
x},\,t)\,d^nx$. Let $\psi_{_j}({\bfm x},\,t)\in{\mathcal F}$ with
$j=1,\ldots,p$ a set of $p$ fields on $M$, with $t$ a real parameter
and we denote by ${\bfm\Omega}\equiv(\psi_{_1},\ldots,\psi_{_p})$ a
$p$-dimensional vector on ${\mathcal M}=M\times\ldots\times M$. We
assume uniform boundary conditions, to guarantee the convergence of
the integrals, by requiring that all fields and their spatial
derivatives vanish quickly on the
boundary of ${\mathcal M}$.\\
In the following we consider a nonrelativistic canonical system
described by the action ${\mathcal A}=\int{\mathcal
L}[{\bfm\Omega}]\,d^nx\,dt$, where ${\mathcal L}[{\bfm\Omega}]\in F$
is the Lagrangian density which depends on the scalar fields
$\psi_{_j}\in M$ and their space and time derivatives. Hereinafter
we use the notation between square brackets $G[\psi]$ to indicate
the dependence of $G$ on the field $\psi$ and its spatial derivative
of any order. Since the theory is nonrelativistic
the Lagrangian contains only time derivatives of the first order.\\
We introduce the variation of a functional $G$ with respect to
$\bfm\Omega$ as
\begin{equation}
\delta G=\left(\frac{\delta G}{\delta\psi_{_1}},\ldots, \frac{\delta
G}{\delta\psi_{_p}}\right) \ ,\label{va}
\end{equation}
where the functional derivative can be defined by means of the Euler
operator \cite{Olver}
\begin{equation}
\frac{\delta G}{\delta\psi_{_j}}\equiv{\mathcal E}_{_{\psi_{_j}}}(G)
\ ,\label{euler}
\end{equation}
given by
\begin{equation}
{\mathcal E}_{_{\psi_{_j}}}(G)=-\frac{\partial}{\partial
t}\left(\frac{\partial{\mathcal
G}[{\bfm\Omega}]}{\partial(\partial_t\psi_{_j})}\right)+\sum_{[k=0]}(-1)^k{\mathcal
D}_{_{I_k}}\left( \frac{\partial{\mathcal
G}[{\bfm\Omega}]}{\partial({\mathcal D}_{_{I_k}}\psi_{_j})}\right) \
,\label{derfun}
\end{equation}
with ${\mathcal D}_{_{I_k}}\equiv\partial^k/(\partial
x_{_1}^{i_1}\ldots
\partial x_{_n}^{i_n})$ and
$\sum_{[k=0]}\equiv\sum_{k=0}^\infty\sum_{I_k}$. The sum
$\sum_{I_k}$ is over the multi-index $I_k\equiv(i_1,\,\ldots,\,i_n)$
with
$0\leq i_q\leq k$, $\sum_q i_q=k$ and $1\leq q\leq n$.\\
It is easy to show, by using equation (\ref{derfun}), that the Euler
operator satisfies the following property
\begin{equation}
{\mathcal E}\left(\frac{\partial B}{\partial
t}+{\bfm\nabla}\cdot{\bfm C}\right)=0 \ ,\label{null}
\end{equation}
where $B$ and the components of ${\bfm C}$ belong to $F$ whilst
${\bfm\nabla}\equiv(\partial_{_1},\,\ldots,\,\partial_{_n})$ is the
$n$-dimensional gradient operator.

\sect{Nonlinear Schr\"odinger equation for scalar particles} We
begin by consider nonrelativistic many body systems of scalar
interacting particles described, in the mean field approximation,
through a very genera family of NLSEs. It is useful to recall the
main aspects of the canonical theory both
in the wave-function formulation and in the hydrodynamic formulation.\\

{\em a) Wave-function formulation.}\\

Let us consider the class of canonical NLSEs described by the
Lagrangian density
\begin{equation}
{\mathcal
L}[\psi^\ast,\,\psi]=\frac{i}{2}\,\left(\psi^\ast\,\frac{\partial\psi}{\partial
t}- \psi\,\frac{\partial\psi^\ast}{\partial t}\right)-
|{\bfm\nabla}\psi|^2-U[\psi^\ast,\,\psi] \ ,\label{lagrangean}
\end{equation}
which is a functional of the 2-dimensional vector
${\bfm\Omega}\equiv(\psi^\ast,\,\psi)$. The first two terms are the
same encountered in the standard linear quantum mechanics whilst the
last term is the nonlinear potential describing the interaction
among the particles of the system. We assume that $U[\psi^\ast,\,\psi]$
be a real
smooth functional of the fields $\psi$ and $\psi^\ast$ and their
spatial derivatives which leave the Lagrangian density
(\ref{lagrangean}) invariant under transformations
belonging to the U(1) group that assures the conservation of the
total number of particles. As we will show, this condition imposes a
constraint on the form of the nonlinear potential
$U[\psi,\,\psi^\ast]$.\\
By introducing the action of the system
\begin{equation}
{\mathcal A}=\int\limits_{\mathcal R}{\mathcal
L}[\psi^\dag,\,\psi]\,d^nx\,dt \ ,\label{action}
\end{equation}
where the domain of integration is the whole real region ${\mathcal
R}=M\times I\!\!R$, the evolution equation for the vector field
$\psi$, corresponding to the stationary trajectories of the action,
can be obtained from the extremal problem
\begin{equation}
\delta{\mathcal A}=0 \ ,\label{vv}
\end{equation}
where the variation is performed with respect to the vector
${\bfm\Omega}$. According to equation (\ref{va}), the Euler-Lagrange
equations for the fields $\psi$ and $\psi^\ast$ are given by
\begin{eqnarray}
\nonumber &&\frac{\delta}{\delta\psi^\ast}\int\limits_{\mathcal R}
\frac{i}{2}\left(\psi^\ast\,\frac{\partial\psi}{\partial t}-
\psi\,\frac{\partial\psi^\ast}{\partial t}\right)\,d^nx\,dt-
\frac{\delta}{\delta\psi^\ast}\int\limits_{\mathcal R}|{\bfm\nabla}\psi|^2\,d^nx\,dt\\
&-&\frac{\delta}{\delta\psi^\ast}\int\limits_{\mathcal R}
U[\psi^\ast,\,\psi]\,d^nx\,dt=0 \ ,
\end{eqnarray}
and its conjugate. We recall that the Lagrangian density
(\ref{lagrangean}) is defined modulo a total derivative (null
Lagrangian) which does not give contribute to the evolution equations because, as
stated in equation (\ref{null}), the variation of a total derivative vanishes.\\
After performing the functional derivatives we obtain the following
NLSE
\begin{equation}
i\,\frac{\partial\psi}{\partial t}+\Delta\psi
+\Lambda[\psi^\ast,\,\psi]=0 \ ,\label{schroedingerr}
\end{equation}
where $\Delta\equiv\partial^2_{_1}+\ldots+\partial^2_{_n}$ is the
$n$ dimensional Laplacian operator and the complex nonlinear term
$\Lambda[\psi^\ast,\,\psi]$ is given by
\begin{equation}
\Lambda[\psi^\ast,\,\psi]=-\frac{\delta}{\delta\psi^\ast}\int\limits_{\mathcal
R} U[\psi^\ast,\,\psi]\,d^nx\,dt \ .
\end{equation}

{\em b) Hydrodynamic formulation.}\\

In the hydrodynamic formulation we introduce two real fields $\rho$
and $S$ related to the wave-function through the polar decomposition
\begin{equation}
\psi({\bfm x},t)=\rho^{1/2}({\bfm x},t)\,\exp\Big(i\,S({\bfm
x},t)\Big) \ ,\label{ansatz}
\end{equation}
or equivalently
\begin{eqnarray}
&&\rho({\bfm x},\,t)=|\psi({\bfm x},\,t)|^2 \ ,\label{r}\\&&\nonumber\\
&&S({\bfm x},\,t)=\frac{i}{2}\,\log\left(\frac{\psi^\ast({\bfm
x},\,t)}{\psi({\bfm x},\,t)}\right) \ .\label{s}
\end{eqnarray}
By defining the action ${\mathcal A}=\int_{\mathcal R}{\mathcal
L}[\rho,\,S]\,d^nx\,dt$ through the Lagrangian density
\begin{equation}
{\mathcal L}[\rho,\,S]=-\frac{\partial S}{\partial t}\,\rho
-({\bfm\nabla} S)^2\,\rho-\frac{({\bfm\nabla}
\rho)^2}{4\,\rho}-U[\rho,\,S] \ ,\label{lagrangiana1}
\end{equation}
where $U[\rho,\,S]$ is the nonlinear potential in the hydrodynamic
representation, from the variational problem $\delta\,{\mathcal
A}=0$, where now ${\bfm\Omega}\equiv(\rho,\,S)$, we obtain two real equations
\begin{eqnarray}
&&\frac{\partial S}{\partial t}+({\bfm\nabla} S)^2+U_q[\rho]-W[\rho,\,S]=0 \ ,\label{hj}\\
&&\frac{\partial\rho}{\partial t}+{\bfm
\nabla}\cdot(2\,\rho\,{\bfm\nabla} S)+2\,\rho\,{\mathcal
W}[\rho,\,S]=0 \ .\label{conti}
\end{eqnarray}
In equations (\ref{hj}) and (\ref{conti})
$U_q[\rho]=-\Delta\rho^{1/2}/\rho^{1/2}$ denotes the quantum
potential \cite{Bohm,Madelung} whilst the two real functionals
$W[\rho,\,S]$ and ${\mathcal W}[\rho,\,S]$ are given by
\begin{eqnarray}
&&W[\rho,\,S]=-\frac{\delta}{\delta\,\rho}\int\limits_{\mathcal R}
U[\rho,\,S]\,d^nx\,dt \ ,\label{real}\\ &&{\mathcal
W}[\rho,\,S]=-\frac{1}{2\,\rho}\frac{\delta}{\delta\,S}\int\limits_{\mathcal
R} U[\rho,\,S]\,d^nx\,dt \ .\label{imaginary}
\end{eqnarray}
According to the relation
\begin{equation}
\frac{\delta}{\delta\psi}=\psi\,\left(\frac{\delta}{\delta\rho}
+\frac{i}{2\,\rho}\,\frac{\delta}{\delta S}\right) \ ,
\end{equation}
the quantities $W[\rho,\,S]$ and ${\mathcal W}[\rho,\,S]$ are
related to the nonlinearity $\Lambda[\psi,\,\psi^\ast]$ in
\begin{equation}
\Lambda[\psi,\,\psi^\ast]=\Big(W[\rho,\,S]+i\,{\mathcal
W}[\rho,\,S]\Big)\,\psi \ .\label{ll}
\end{equation}
Equation (\ref{hj}) is an {\it Hamilton-Jacobi}-like equation for
the field $S$ whilst equation (\ref{conti}) describes the evolution
equation for the field $\rho$. The last term in equation
(\ref{conti}) originates from the nonlinear potential $U[\rho,\,S]$
and is a source of particle which, in the general case, destroys the
conservation of the number of particles of the system.\\
Accounting for equation (\ref{ll}) we can rewrite the NLSE
(\ref{schroedingerr}) in
\begin{equation}
i\,\frac{\partial\psi}{\partial t}+\Delta\psi
+\Big(W[\rho,\,S]+i\,{\mathcal W}[\rho,\,S]\Big)\,\psi=0 \
,\label{schroedinger1}
\end{equation}
where the complex nonlinearity $\Lambda[\rho,\,S]$, expressed in the
hydrodynamic fields $\rho$ and $S$, is separated in the real
$W[\rho,\,S]$ and imaginary ${\mathcal W}[\rho,\,S]$ part.


\subsection{U(1) Symmetry} Differently from the linear Schr\"odinger
equation which is U(1)-invariant, the presence of the nonlinearity
$U[\rho,\,S]$ generally breaks this symmetry. In fact, equation
(\ref{conti}), in general, is not a continuity equation for the
field $\rho$. In the following we study the relevant restrictions to
the nonlinear potential $U[\rho,\,S]$ so that the Lagrangian
(\ref{lagrangean}) turns out to be U(1)-invariant. Such invariance
implies, according to the Noether theorem \cite{Noether}, the
conservation of the total number of particles by restoring
a continuity equation for the field $\rho$. \\
To begin with, we rewrite equation (\ref{conti}) in the form
\begin{equation}
\frac{\partial\rho}{\partial t}+{\bfm\nabla}\cdot{\bfm
j}^{(0)}_{_\psi}+2\,\rho\,{\mathcal W}[\rho,\,S]=0 \ ,\label{eq2}
\end{equation}
where
\begin{equation}
{\bfm j}^{(0)}_{_\psi}=-i\,\Big(\psi^\ast\,{\bfm\nabla}\psi
-\psi\,{\bfm\nabla}\psi^\ast\Big)\equiv2\,\rho\,{\bfm\nabla}S \
,\label{currentcct}
\end{equation}
is the bilinear particles current of the standard quantum mechanics.
Making use of equation (\ref{imaginary}), equation (\ref{eq2})
become
\begin{equation}
\frac{\partial\rho}{\partial t}+{\bfm\nabla}\cdot{\bfm
j}^{(0)}_{_\psi}-\,\frac{\delta}{\delta S}\int\limits_{\mathcal R}
U[\rho,\,S]\,d^nx\,dt=0 \ ,\label{bilin}
\end{equation}
and taking into account the definition of functional derivative
(\ref{euler}) we get
\begin{eqnarray}
\nonumber \frac{\partial\rho}{\partial t}&+&{\bfm\nabla}\cdot{\bfm
j}^{(0)}_{_\psi}-\frac{\partial}{\partial S}\int\limits_{\mathcal R}
U[\rho,\,S]\,d^nx\,dt\\
&-&\sum_{[k=1]}(-1)^k{\mathcal
D}_{_{I_k}}\left(\frac{\partial}{\partial({\mathcal
D}_{_{I_k}}S)}\int\limits_{\mathcal R}
U[\rho,\,S]\,d^nx\,dt\right)=0 \ .\label{cuc}
\end{eqnarray}
This last equation can be written in the form
\begin{equation}
\frac{\partial\rho}{\partial t}+{\bfm\nabla}\cdot{\bfm
j}_{_\psi}=\frac{\partial U}{\partial S} \ ,\label{continuityro}
\end{equation}
where the nonlinear current $\bfm j_{_\psi}$ has components
\begin{equation}
\Big({\bfm
j}_{_\psi}\Big)_i=2\,\rho\,\partial_iS+\frac{\delta}{\delta(\partial_iS)}\int\limits_{\mathcal
R} U[\rho,\,S]\,d^nx\,dt \ ,\label{ccc}
\end{equation}
because, according to the definition (\ref{derfun}), the following
relation holds
\begin{eqnarray}
\nonumber & &\sum_{[k=1]}(-1)^k{\mathcal
D}_{_{I_k}}\left(\frac{\partial}{\partial({\mathcal
D}_{_{I_k}}S)}\int\limits_{\mathcal R}
U[\rho,\,S]\,d^nx\,dt\right)\\
&=&-{\bfm\nabla}\cdot
\Bigg(\frac{\delta}{\delta({\bfm\nabla}S)}\,\int\limits_{\mathcal R}
U[\rho,\,S]\,d^nx\,dt\Bigg) \ .
\end{eqnarray}
Remark that the expression of current (\ref{ccc}) is always defined
modulo the curl of an arbitrary functional $G[\rho,\,S]$ which does
not
give contribute to equation (\ref{continuityro}).\\
As stated before, the Lagrangian (\ref{lagrangean}), for a general
nonlinear potential $U[\rho,\,S]$, is not U(1)-invariant. In fact,
equation (\ref{continuityro}) describe a very general kinetics
process in which the right hand side plays the role of a source of
particles. Trivially, if the nonlinear potential $U[\rho,\,S]$
depends on the field $S$ only through its spatial derivatives the
right hand side of equation (\ref{continuityro}) vanishes and it
becomes a continuity equation for the field $\rho$
\begin{equation}
\frac{\partial\rho}{\partial t}+{\bfm\nabla}\cdot{\bfm j}_{_\psi}=0
\ .\label{continu}
\end{equation}
The conserved quantity associated to equation (\ref{continu}) is the
total number of particles
\begin{equation}
N=\int\limits_M\rho\,d^nx \ ,\label{numero}
\end{equation}
where the integral is evaluated on the full real region $M$ (uniform
boundary conditions
guarantee the convergence of the integral). \\
Under the assumption that $U[\rho,\,S]$ depends on $S$ only through
its derivatives, according to the relation
\begin{equation}
\frac{\delta}{\delta S}\int\limits_{\mathcal R}
U[\rho,\,S]\,d^nx\,dt=-{\bfm\nabla}\cdot\left(\frac{\delta}{\delta({\bfm\nabla}S)}\int\limits_{\mathcal
R} U[\rho,\,S]\,d^nx\,dt\right) \ ,
\end{equation}
we can rewrite the expression of ${\mathcal W}[\rho,\,S]$ in
\begin{equation}
{\mathcal W}[\rho,\,S]=\frac{1}{2\,\rho}\,{\bfm\nabla}\cdot{\bfm
{\mathcal J}}[\rho,\,S] \ ,\label{funf1}
\end{equation}
where $\bfm{\mathcal J}[\rho,\,S]$ is given by
\begin{equation}
\left(\bfm{\mathcal
J}\right)_i[\rho,\,S]=\frac{\delta}{\delta(\partial_iS)}\int\limits_{\mathcal
R} U[\rho,\,S]\,d^nx\,dt \ ,\label{funf}
\end{equation}
and the expression of the current (\ref{ccc}) becomes
\begin{equation}
{\bfm j}_{_\psi}={\bfm j}_{_\psi}^{(0)}+{\bfm{\mathcal J}
}[\rho,\,S] \ .\label{cccc}
\end{equation}
Summing up, the nonlinear potential $U[\rho,\,S]$ generally breaks
the U(1) symmetry of the system (\ref{lagrangean}). On the other
hand, if it depends on the field $S$ only
through its spatial derivative, the U(1) symmetry is restored. This
can be clarified if we taking into account that, under a global U(1)
transformation (gauge transformation of first kind)
\begin{equation}
\psi\rightarrow\phi=e^{i\,\epsilon}\,\psi \ ,\label{u1}
\end{equation}
the phase $S$ transforms in
\begin{equation}
S\rightarrow {\mathcal S}=S+\epsilon \ ,
\end{equation}
where $\epsilon$ is the constant parameter of the transformation. As
a consequence, if $S$ appears in the Lagrangian only through its
derivatives, the transformation (\ref{u1}) does not change the
Lagrangian of the system.


\subsection{Gauge transformation}

We introduce a unitary and nonlinear transformation on the field
$\psi$
\begin{equation}
\psi({\bfm x},\,t)\rightarrow\phi({\bfm x},\,t)={\mathcal
U}[\rho,\,S]\,\psi({\bfm x},\,t) \ ,\label{transf1}
\end{equation}
whose purpose is to change the NLSE (\ref{schroedinger1}), which
contains a complex nonlinearity $W[\rho,\,S]+i{\mathcal
W}[\rho,\,S]$, in another one containing only a purely real
nonlinearity $\widetilde W[\rho,\,{\mathcal S}]$. As a consequence
the current $\bfm j_{_\psi}$, given in equation (\ref{cccc}), is
transformed in another one ${\bfm j}_{_\psi}\rightarrow{\bfm
j}_{_\phi}^{(0)}$ having the merely bilinear form of the ordinary quantum
mechanics.\\ Since the transformation is unitary: ${\mathcal
U}^\ast={\mathcal U}^{-1}$, equation (\ref{transf1}) does not change
the quantity
\begin{equation}
\rho({\bfm x},\,t)=|\psi({\bfm x},\,t)|^2=|\phi({\bfm x},\,t)|^2 \ ,
\end{equation}
representing the density of probability of position of the system.\\
The functional $\mathcal U[\rho,\,S]$ is defined in
\begin{equation}
{\mathcal U}[\rho,\,S]=\exp\Big(i\,\sigma[\rho,\,S]\Big) \
,\label{trasf2}
\end{equation}
where the generator $\sigma[\rho,\,S]$ is a real functional which
relates the phase $\mathcal S$ of the field $\phi$ with the phase
$S$ of the field $\psi$
\begin{equation}
{\mathcal S}=S+\sigma[\rho,\,S] \ ,\label{phases}
\end{equation}
since we define
\begin{equation}
\phi({\bfm x},\,t)=\rho^{1/2}({\bfm x},\,t)\,\exp\Big(i\,{\mathcal
S}({\bfm x},\,t)\Big) \ .
\end{equation}
When equation (\ref{phases}) is invertible, we can express the phase
$S$ as a functional of the fields $\mathcal S$ and $\rho$.\\
The expression of the generator $\sigma[\rho,\,S]$ is related to the
imaginary part $\mathcal W[\rho,\,S]$ of the NLSE through
\begin{equation}
{\bfm\nabla}\sigma[\rho,\,S]=\frac{1}{2\,\rho}\,{\bfm{\mathcal J}
}[\rho,\,S] \ ,\label{genf}
\end{equation}
which defines $\sigma[\rho,\,S]$ modulo an arbitrary integration
constant. The same equation (\ref{genf}) imposes a condition on the
form of the nonlinear potential as it follows from the relation
${\bfm\nabla}\times{\bfm\nabla}\,\sigma=0$ (where
${\bfm\nabla}\times{\bfm f}$ means
$\partial_{_i}\,f_{_j}-\partial_{_j}\,f_{_i}$ with
$i,\,j=1,\ldots,\,n$)
\begin{equation}
{\bfm\nabla}\times\left(\frac{\bfm{\mathcal
J}[\rho,\,S]}{\rho}\right)=0 \ .\label{rot}
\end{equation}
Equation (\ref{rot}) selects the potentials $U[\rho,\,S]$ and in
this way the nonlinear systems in which we can perform the
transformation (\ref{transf1}). For one-dimensional systems this
transformation can be always accomplished.

By plunging the expression of $\psi({\bfm x},\,t)={\mathcal
U}^{-1}[\rho,\,S]\,\phi({\bfm x},\,t)$ in the NLSE
(\ref{schroedinger1}) it is easy to verify that it reduces in the
following evolution equation
\begin{equation}
i\,\frac{\partial\phi}{\partial t}+\Delta\phi
+\widetilde{W}[\rho,\,{\mathcal S}]\,\phi=0 \ ,\label{schroedinger3}
\end{equation}
which contains only a real nonlinearity
$\widetilde{W}[\rho,\,{\mathcal S}]$ given by
\begin{equation}
\widetilde{W}[\rho,\,{\mathcal
S}]=W-({\bfm\nabla}\,\sigma)^2+2\,{\bfm \nabla}{\mathcal S}
\cdot{\bfm\nabla}\sigma+\frac{\partial\sigma}{\partial t} \
,\label{real1}
\end{equation}
where $W\equiv W[\rho,\,S[\rho,\,{\mathcal S}]]$ and $\sigma\equiv
\sigma[\rho,\,S[\rho,\,{\mathcal S}]]$. Because the phase $\mathcal
S$ appears in equation (\ref{schroedinger3}) only trough its spatial
derivatives, as required by the U(1)-invariance, the arbitrary
integration constant arising from the definition of
${\mathcal\sigma}[\rho,\,S]$, does not produce any effect and can be
posed equal to zero. Finally, the last term in equation
(\ref{real1}) can be solved using the Hamilton-Jacobi equation
(\ref{hj}) and the continuity equation (\ref{conti}) in order to
reduce the nonlinearity $\widetilde{\mathcal W}[\rho,\,S]$ in a
quantity containing only space
derivatives.\\
We can easily verify that the continuity equation for the field
$\rho$, obtained from equation (\ref{schroedinger3}), is now given
by
\begin{equation}
\frac{\partial\rho}{\partial t}+{\bfm\nabla}\cdot{\bfm
j}_{_\phi}^{(0)}=0 \ ,
\end{equation}
where, due to the reality of the nonlinearity $\widetilde{\mathcal
W}[\rho,\,S]$, the current
\begin{equation}
{\bfm j}_{_\phi}^{(0)}=2\,\rho\,{\bfm\nabla}{\mathcal S} \
,\label{corcor}
\end{equation}
assumes the standard bilinear form.\\
Let us briefly discuss the generalization of this transformation to
non canonical systems. Firstly, we observe that for a non canonical
system the two quantities $W[\rho,\, S]$ and ${\mathcal
W}[\rho,\,S]$ are not derivable from a potential $U[\rho,\,S]$. In
particular, the real nonlinearity $W[\rho,\,S]$ can assume any
arbitrary expression whereas the form of the imaginary nonlinearity
${\mathcal W}[\rho,\,S]$, constrained by the
continuity equation for the field $\rho$, is given by
\begin{equation}
{\mathcal
W}[\rho,\,S]={1\over2\,\rho}\,{\bfm\nabla}\cdot\bfm{\mathcal
J}[\rho,\,S] \ ,\label{w}
\end{equation}
for an arbitrary functional $\bfm{\mathcal J}[\rho,\,S]$. The
particle current ${\bfm j}_{_\psi}$ is still given through the
relation
\begin{equation}
{\bfm j}_{_\psi}=2\,\rho\,{\bfm\nabla}S+\bfm{\mathcal J}[\rho,\,S] \
,\label{noncan}
\end{equation}
but now the functional $\bfm{\mathcal J}[\rho,\,S]$ is related to
${\mathcal W}[\rho,\,S]$ only through equation (\ref{w}). Following
the same steps described for the canonical case, it is easy to
verify that the transformation (\ref{transf1}) with generator
defined in the same way
\begin{equation}
{\bfm\nabla}\sigma[\rho,\,S]={1\over2\,\rho}\,\bfm{\mathcal
J}[\rho,\,S] \ ,
\end{equation}
eliminates the imaginary part of the nonlinearity in the motion
equation (\ref{schroedinger1}) which assumes the expression
(\ref{schroedinger3})-(\ref{real1}). We observe that, differently
from the canonical case, equation (\ref{rot}) now constraints only
the form of the imaginary part ${\mathcal W}[\rho,\,S]$ whilst the
real part $W[\rho,\,S]$ is completely arbitrary.\\
Finally, let us remark that when the transformation (\ref{transf1})
is applied to a canonical equation, generally it breaks the
canonical structure of the theory. Consequently, the new NLSE is no
more expressible in the Lagrangian formalism. Differently, when the
transformation is applied to a non canonical system, the new NLSE
can acquire a canonical structure. It is not hard to show that this
is possible if the transformed nonlinearity $\widetilde W[\rho]$ is
a functional depending only on the field $\rho$. In fact, the real
quantity $\widetilde W$ is related to a nonlinear potential
$\widetilde U$ through equation (\ref{real}) and because the new
nonlinearity is purely real, from equation (\ref{imaginary}) it
follows that the potential $\widetilde U$ and consequently
$\widetilde W$ cannot depend on the field $\mathcal S$.

\sect{Coupled nonlinear Schr\"odinger equations}
Physical systems, whose dynamics is described by means of CNLSEs are
ubiquitous in nature. They occur, for instance, in presence of many
interacting particle of different spices or in presence of
multi polarized laser beams propagating in optical fibers. In this last case, any polarized component of the electric or
magnetic field, can be considered like a ``particle state'' since
its evolution is describable through a NLS-like equation \cite{Hasegawa}.\\
We observe that for many particle systems of different kind, by
denoting with $N_{_k}$ the number of the $k$th species, many
possible combinations of conserved multiplets can be realized.
In particular, two relevant physical situations are given when:\\
a) All the quantities $N_{_k}$ are separately conserved, which is
relevant, for instance, in nonrelativistic systems of multi-species
where process of transmutation from a species
to another one is forbidden.\\
b) Only the quantity $N_{\rm tot}=\sum_kN_{_k}$ is conserved.
Relevant examples are given in the study of light propagation in
optical fibers where each species describes a polarization mode and
only the total intensity of the beam is conserved.\\
In the following we introduce a wide class of CNLSEs in the form
\begin{equation}
i\,\frac{\partial\Psi}{\partial t}+\widehat{A}\,\Delta
\Psi+\widehat\Lambda[{\vec \rho},\,{\vec S}]\,\Psi=0 \
,\label{csch1}
\end{equation}
where $\Psi=(\psi_{_1},\,\ldots,\,\psi_{_p})$,
${\vec\rho}\equiv(\rho_{_1}\,\ldots,\,\rho_{_p})$ and ${\vec
S}\equiv(S_{_1},\,\ldots,\,S_{_p})$ are $p$ dimensional vectors.
We denote the operator valued matrix $\widehat{M}[{\vec
v}]$ by an hat (the lower case letter $m[{\vec v}]$ denotes its
entries) and use the notation between square brackets to indicate
the functional dependence on the components of the vector ${\vec
v}=(v_{_1},\,\ldots,\,v_{_p})$ and on its spatial derivatives of any
order. Without loss of generality we assume the $p\times p$ matrix
$\widehat A$ in a diagonal
form.\\
We observe that any system of CNLSEs can be always accommodated in
the form given in equation (\ref{csch1}) with a diagonal
nonlinearity $\widehat\Lambda[{\vec \rho},\,{\vec S}]$. Such
nonlinearity can be separated in an Hermitian matrix $\widehat
W=(\widehat\Lambda+\widehat\Lambda^\dag)/2$ and an anti-Hermitian
matrix $i\,\widehat{\mathcal
W}=(\widehat\Lambda-\widehat\Lambda^\dag)/2$. Thus, we can pose
$\widehat\Lambda[{\vec \rho},\,{\vec S}]=\widehat W[{\vec
\rho},\,{\vec S}]+i\,\widehat{\mathcal W}[{\vec \rho},\,{\vec S}]$,
where the diagonal matrices $\widehat W[{\vec \rho},\,{\vec S}]$ and
$\widehat{\mathcal W}[{\vec \rho},\,{\vec S}]$ have purely real
entries. Such assumption is only for sake of convenience
and does not imply any restriction on the form of the nonlinearity.\\
We will consider a general situation in which the system
(\ref{csch1}) has $q$ conserved multiplets of order $p_{_k}$, with
$k=1,\,\ldots,\,q$ and $\sum_kp_{_k}=p$, where $1\leq q\leq p$. In
this way, the two particular cases a) and b) quoted previously are
recognized for $q=p$ and $q=1$,
respectively.\\
Let us organize the fields $\psi_{_i}$, belonging to the vector
$\Psi$, in
\begin{equation}
\stackrel{\mbox{$\Psi\equiv$}\big(\underbrace{\psi_{_{11}},
\,\ldots,\, \psi_{_{1p_{_1}}}}\mbox{$;\,$}
\underbrace{\psi_{_{21}},\,\ldots,\,
\psi_{_{2p_{_2}}}}\mbox{$;\,\ldots;\,$} \underbrace{\psi_{_{q1}},\,
\ldots,\,\psi_{_{qp_{_q}}}}\big)\mbox{$ \ ,$}}
{\scriptscriptstyle\hspace{6mm} {\rm 1st\;
multiplet\hspace{10mm}2nd\; multiplet\hspace{16mm}}q{\rm th\;
multiplet}}
\end{equation}
and, from now on, we relabel the fields $\psi_{_i}$ in
$\psi_{_{kl}}$ where, the first index $k$ refers to the $k$th
multiplet of order $p_{_k}$, whereas the second index $l$, with
$1\leq l\leq p_{_k}$, refers to the $l$th field inside to the
multiplet $k$.\\
Canonical system (\ref{csch1}) can be obtain from the Lagrangian
density
\begin{equation}
{\mathcal L}[\Psi^\dag,\,\Psi]=\frac{i}{2}\,\left(\Psi^\dag\,
\frac{\partial\Psi}{\partial t}-\frac{\partial\Psi^\dag}{\partial t}
\,\Psi\right)-{\bfm\nabla}\Psi^\dag\cdot\widehat
A\,{\bfm\nabla}\Psi-U[\Psi^\dag,\,\Psi] \ ,\label{qlagrangean}
\end{equation}
where the scalar product in the second term is applied among the
gradient operators. The nonlinear potential $U[\Psi^\dag,\,\Psi]$ is
a real functional which depends on the vector fields $\Psi$,
$\Psi^\dag$ and their spatial derivatives. Accounting for the
uniform boundary conditions, the potential $U[\Psi^\dag,\,\Psi]$
vanishes together with all its derivatives, at the spatial
infinity. \\
By introducing the action of the system
\begin{equation}
{\mathcal A}=\int\limits_{\mathcal R}{\mathcal
L}[\Psi^\dag,\,\Psi]\,d^nx\,dt \ ,\label{aact}
\end{equation}
the evolution equation for the vector field $\Psi$ is given by the
stationary trajectories of the action as it follows from the
variational problem $\delta{\mathcal A}=0$, where the variation is
performed with respect to the $2p$-dimensional vector
${\bfm\Omega}\equiv(\Psi^\dag,\,\Psi)$.\\ In this way we obtain the
equation
\begin{equation}
i\,\frac{\partial\Psi}{\partial t}+\widehat{A}\,\Delta
\Psi-\frac{\delta}{\delta\Psi^\dag}\int\limits_{\mathcal R}
U[\Psi^\dag,\,\Psi]\,d^nx\,dt=0 \ ,\label{c1NLSE}
\end{equation}
and its Hermitian conjugate, which form a system of $2p$-nonlinear
coupled Schr\"odinger equations.\\
Taking in account of the polar decomposition of the fields
$\psi_{_{kl}}$ in the real fields $\rho_{_{kl}}$ and $S_{_{kl}}$
\begin{equation}
\psi_{_{kl}}({\bfm x},\,t)=\rho_{_{kl}}^{1/2}({\bfm x},\,t)
\,\exp\Big(i\,S_{_{kl}}({\bfm x},\,t)\Big) \ ,\label{hydro}
\end{equation}
we can express the variation $\delta/\delta\psi_{_{kl}}^\ast$ as
\begin{equation}
\frac{\delta}{\delta\psi_{_{kl}}^\ast}=
\psi_{_{kl}}\,\left(\frac{\delta}{\delta\rho_{_{kl}}}
+\frac{i}{2\,\rho_{_{kl}}}\,\frac{\delta} {\delta S_{_{kl}}}\right)
\ .
\end{equation}
In this way, each component of equation (\ref{c1NLSE}) can be
written in
\begin{equation}
i\,\frac{\partial\psi_{_{kl}}}{\partial
t}+a_{_{kl}}\,\Delta\psi_{_{kl}}-\left[\left(\frac{\delta}{\delta
\rho_{_{kl}}} +\frac{i}{2\,\rho_{_{kl}}}\,\frac{\delta} {\delta
S_{_{kl}}}\right)\int\limits_{\mathcal R} U[{\vec\rho},\,{\vec
S}]\,d^nx\,dt\right]\,\psi_{_{kl}}=0 \ , \label{eqqq}
\end{equation}
where $U[{\vec\rho},\,{\vec S}]$ is the nonlinear potential in the
hydrodynamic representation. Equation (\ref{eqqq}) can be posed in
the following matrix form
\begin{equation}
i\,\frac{\partial\Psi}{\partial t}+\widehat{A}\,\Delta\Psi
+\left(\widehat W[{\vec\rho},\,{\vec S}]+i\,\widehat {\mathcal
W}[{\vec\rho},\,{\vec S}]\right)\,\Psi=0 \ ,\label{c2NLSE}
\end{equation}
where the Hermitian and anti-Hermitian nonlinearities are given by
\begin{eqnarray}
&&\widehat W[{\vec\rho},\,{\vec S}]=-{\rm
diag}\left(\frac{\delta}{\delta\rho_{_{kl}}}\int\limits_{\mathcal
R}U[{\vec\rho},\,{\vec S}]\,dx\,dt\right) \ ,\label{hermit}\\
&&\widehat{\mathcal W}[{\vec\rho},\,{\vec S}]=-{\rm
diag}\left(\frac{1}{2\,\rho_{_{kl}}} \,\frac{\delta}{\delta
S_{_{kl}}}\int\limits_{\mathcal R} U[{\vec\rho},\,{\vec
S}]\,dx\,dt\right)
 \ .\label{antihermitt}
\end{eqnarray}
Finally, by using the polar decomposition (\ref{hydro}), equation
(\ref{c2NLSE}) can be separate in a system of $2p$ nonlinear real
coupled equations
\begin{eqnarray}
&&\frac{\partial S_{_{kl}}}{\partial\,t}+a_{_{kl}}
\,({\bfm\nabla}S_{_{kl}})^2- a_{_{kl}}\,\frac{\Delta
\rho_{_{kl}}^{1/2}}{\rho_{_{kl}}^{1/2}}-w_{_{kl}}
[{\vec\rho},\,{\vec S}]=0 \
,\label{ms}\\&&\frac{\partial\rho_{_{kl}}}{\partial t}
+2\,a_{_{kl}}\,{\bfm\nabla}\cdot( \rho_{_{kl}}\,{\bfm\nabla}
S_{_{kl}}) +2\,\rho_{_{kl}}\,{\scriptstyle{\mathcal
W}}_{_{kl}}[{\vec\rho},\,{\vec S}]=0 \ .\label{mr}
\end{eqnarray}
The first set of equations (\ref{ms}) is a system of $p$-coupled
{\em Hamilton-Jacobi}-like equations for the fields $S_{_{kl}}$,
whilst the second set of equations (\ref{mr}) describes the time
evolution of the fields $\rho_{_{kl}}$.

\subsection{U(1) symmetry}

In the following we consider those systems written in the form
(\ref{c2NLSE}) which admit a set of $q$ continuity equations
\begin{equation}
\frac{\partial\rho_{_{k}}}{\partial t}+{\bfm\nabla}\cdot{\bfm
j}_{_{\Psi,k}}=0 \ ,\label{qcon}
\end{equation}
which assure the conservation of the quantities
\begin{equation}
N_{_k}=\int\limits_M\rho_{_k}\,d^nx \ .\label{ntot}
\end{equation}
This imposes some restrictions on the functional dependence of the
potential $U[{\vec\rho},\,{\vec S}]$ with respect to the fields
$\vec\rho$ and $\vec S$. To obtain such restrictions we recall the
following relation
\begin{equation}
\frac{\delta}{\delta S_{_{kl}}}=\frac{\partial} {\partial
S_{_{kl}}}-
{\bfm\nabla}\cdot\frac{\delta}{\delta({\bfm\nabla}S_{_{kl}})} \
,\label{derfun1}
\end{equation}
so that, by taking in account the expression of the matrix $\widehat
{\mathcal W}[\vec \rho,\,\vec S]$, we can rewrite equation
(\ref{mr}) in
\begin{eqnarray}
\nonumber \frac{\partial\rho_{_{kl}}}{\partial t}
&+&{\bfm\nabla}\cdot\left(2\,a_{_{kl}}
\,\rho_{_{kl}}\,{\bfm\nabla}S_{_{kl}}+
\frac{\delta}{\delta({\bfm\nabla}S_{_{kl}})} \int\limits_{\mathcal
R} U[{\vec\rho},\,{\vec S}]\,d^nx\,dt\right)\\&-&\frac{\partial}
{\partial S_{_{kl}}}\int\limits_{\mathcal R} U[{\vec\rho},\,{\vec
S}]\,d^nx\,dt=0 \ .
\end{eqnarray}
By summing this equation on the index $l$, with $1\leq l\leq
p_{_k}$, we obtain
\begin{equation}
\frac{\partial\rho_{_k}}{\partial t}+{\bfm\nabla}\cdot\Big({\bfm
j}^{(0)}_{_{\Psi,k}}+\bfm{\mathcal J}_{_k}[{\vec\rho},\,{\vec
S}]\Big)+ I_{_k}[{\vec\rho},\,{\vec S}]=0 \ ,\label{ccsor1}
\end{equation}
where
\begin{equation}
\rho_{_k}=\sum_{l=1}^{p_{_k}}\rho_{_{kl}} \ ,
\end{equation}
and
\begin{equation}
{\bfm j}^{(0)}_{_{\Psi,k}}=\sum_{l=1}^{p_k}{\bfm
j}_{_{\Psi,kl}}^{(0)} \ ,
\end{equation}
with
\begin{equation}
{\bfm j}^{(0)}_{_{\Psi,kl}}=2\,a_{_{kl}}
\,\rho_{_{kl}}\,{\bfm\nabla}S_{_{kl}} \ .\label{j0}
\end{equation}
Moreover, we posed
\begin{equation}
\bfm{\mathcal J}_{_k}[{\vec\rho},\,{\vec
S}]=\sum_{l=1}^{p_k}\bfm{\mathcal J}_{_{kl}}[{\vec\rho},\,{\vec S}]
\ ,\label{jjc}
\end{equation}
with
\begin{equation}
\left(\bfm{\mathcal J}_{_{kl}}\right)_i[{\vec\rho},\,{\vec
S}]=\frac{\delta}{\delta(\partial_iS_{_{kl}})} \int\limits_{\mathcal
R} U[{\vec\rho},\,{\vec S}]\,d^nx\,dt \ ,\label{antientry}
\end{equation}
whilst
\begin{equation}
I_{_k}[{\vec\rho},\,{\vec S}]=\sum_{l=1}^{p_k}
I_{_{kl}}[{\vec\rho},\,{\vec S}] \ ,\label{iic}
\end{equation}
with
\begin{equation}
I_{_{kl}}=-\frac{\partial} {\partial S_{_{kl}}}
\int\limits_{\mathcal R} U[{\vec\rho},\,{\vec S}]\,d^nx\,dt \
.\label{ic}
\end{equation}
By comparing equation (\ref{qcon}) with equation (\ref{ccsor1}) we
obtain, as a condition, that the functionals
$I_{_k}[{\vec\rho},\,{\vec S}]$ must be expressed as the gradient of
a set of functionals ${\bfm G}_{_k}[{\vec\rho},\,{\vec S}]$
\begin{equation}
I_{_k}[{\vec\rho},\,{\vec S}]={\bfm\nabla}\cdot{\bfm
G}_{_{k}}[{\vec\rho},\,{\vec S}] \ .\label{cond}
\end{equation}
We remark that the expression of the functionals ${\bfm
G}_{_k}[{\vec\rho},\,{\vec S}]$ is determined univocally from the
nonlinear potential $U[{\vec\rho},\,{\vec S}]$ through equations
(\ref{iic}), (\ref{ic}) and (\ref{cond}) which select, in this way,
the class of the Lagrangians (\ref{qlagrangean}) of the family of
CNLSEs compatible with the set of continuity equations (\ref{qcon}).
If the conditions (\ref{cond}) are accomplished, equations
(\ref{ccsor1}) become a system of $q$ continuity equations, where
the total currents of the $k$th multiplet ${\bfm j}_{_{\Psi,k}}$ is
given by
\begin{equation}
{\bfm j}_{_{\Psi,k}}={\bfm j}^{(0)}_{_{\Psi,k}}+\bfm{\mathcal
J}_{_k}[{\vec\rho},\,{\vec S}]+{\bfm G}_{_k}[{\vec\rho},\,{\vec S}]
\ .\label{ccurrent}
\end{equation}
We recall that, as it follows from the Noether theorem, equations
(\ref{qcon}) are consequence of the invariance of the Lagrangian
with respect to a global unitary transformation
\begin{equation}
\Psi\rightarrow\Phi=\widehat U\,\Psi \ ,\label{tr1}
\end{equation}
where
\begin{equation}
\widehat U={\rm diag}\Big(\exp(i\,{\vec\epsilon})\Big) \ ,
\end{equation}
and
\begin{equation}
\stackrel{\mbox{${\vec\epsilon}\equiv$}\big(\underbrace{\epsilon_{_1},\,\ldots,\,
\epsilon_{_1}}\mbox{$;\,$}\underbrace{\epsilon_{_2},\,\ldots,\,
\epsilon_{_2}}\mbox{$;\,\ldots;\,$}\underbrace{\epsilon_{_q},\,
\ldots,\,\epsilon_{_q}}\big)\mbox{$ \ ,$}} {\scriptstyle\hspace{4mm}
p_{_1}{\rm \;times}\hspace{7mm}p_{_2}{\rm
\;times}\hspace{14mm}p_{_q}{\rm\; times}}
\end{equation}
are the constant parameters of the transformation.\\
In fact, the Lagrangian (\ref{qlagrangean}) is invariant under the
transformation (\ref{tr1}) if the nonlinear potential
$U[{\vec\rho},\,{\vec S}]$ changes according to
\begin{equation}
\delta\,U[{\vec\rho},\,{\vec
S}]=-\sum_{k=1}^q\epsilon_{_k}\,{\bfm\nabla}\cdot{\bfm
G}_{_{k}}[{\vec\rho},\,{\vec S}] \ ,\label{ttrr1}
\end{equation}
where ${\bfm G}_{_k}[{\vec\rho},\,{\vec S}]$ are arbitrary
functionals. We recall that in this way the motion equation
(\ref{c1NLSE}) does not change because the Lagrangian density is
always defined modulo a total derivative of an arbitrary functional.
Taking in account for the independence of the parameters
$\epsilon_{_k}$, from equation (\ref{ttrr1}) we obtain
\begin{equation}
\sum_{l=1}^{p_{_k}}\frac{\partial}{\partial S_{_{kl}}} \int
U[{\vec\rho},\,{\vec S}]\,d^nx\,dt =-{\bfm\nabla}\cdot{\bfm
G}_{_{k}}[{\vec\rho},\,{\vec S}] \ ,\label{condd1}
\end{equation}
which, according to the definitions (\ref{iic}) and (\ref{ic})
coincides with the condition (\ref{cond}). In addition, because the
parameters $\epsilon_{_k}$ are constants, the potential
$U[{\vec\rho},\,{\vec S}]$ can depend on the phases $S_{_{kl}}$ also
through their spatial derivatives of any order.


\subsection{Gauge transformation}

We are ready to generalize the nonlinear gauge transformation
described in the section 3.2 to the family of CNLSEs under
inspection. Let us introduce the following transformation
\begin{equation}
\Psi({\bfm x},\,t)\rightarrow\Phi({\bfm x},\,t)=\widehat{\mathcal
U}[{\vec\rho},\,{\vec S}]\,\Psi({\bfm x},\,t) \ ,\label{ctr}
\end{equation}
where $\widehat{\mathcal U}[\vec\rho,\,\vec S]$ is a diagonal and
unitary matrix: $ \widehat{\mathcal U}^\dag=\widehat{\mathcal
U}^{-1}$. This implies
\begin{equation}
\rho_{_{kl}}({\bfm x},\,t)=\left|\,\psi_{_{kl}}({\bfm
x},\,t)\right|^2= \left|\,\phi_{_{kl}}({\bfm x},\,t)\right|^2  \ ,
\end{equation}
whilst the phases ${\mathcal S}_{_{kl}}$ are related to fields
$\phi_{_{kl}}$ through the relation
\begin{equation}
{\mathcal
S}_{_{kl}}=\frac{i}{2}\,\ln\left(\frac{\phi_{_{kl}}^\ast({\bfm
x},\,t)}{\phi_{_{kl}}({\bfm x},\,t)}\right) \ ,
\end{equation}
since we define
\begin{equation}
\phi_{_{kl}}({\bfm x},\,t)=\rho_{_{kl}}^{1/2}({\bfm
x},\,t)\,\exp\Big(i\,{\mathcal S}({\bfm x},\,t)\Big) \ .
\end{equation}
Without lost of generality, the matrix $\widehat{\mathcal
U}[\vec\rho,\,\vec S]$ can be written in
\begin{equation}
\widehat{\mathcal U}[{\vec\rho},\,{\vec S}]={\rm diag}\,\Big(
\exp\Big(i\,{\vec \sigma}[{\vec\rho},\,{\vec S}]\Big)\Big) \
,\label{u11}
\end{equation}
where ${\vec\sigma} \equiv(\ldots,\,\sigma_{_{kl}},\,\ldots)$ is a
$p$-dimensional vector with real components. The generators of the
transformation $\sigma_{_{kl}}[{\vec\rho},\,{\vec S}]$ relate the
phase $\vec{\mathcal S}$ of the new field $\Phi$ with the phase
${\vec S}$ of the old field $\Psi$ according to
\begin{equation}
\vec{\mathcal S}={\vec S}+{\vec \sigma}[{\vec\rho},\,{\vec S}] \
.\label{phase}
\end{equation}
We introduce the generators $\sigma_{_{kl}}[{\vec\rho},\,{\vec S}]$
through the relations
\begin{equation}
{\bfm\nabla}\sigma_{_{kl}}[{\vec\rho},\,{\vec S}]=\frac{1}
{2\,a_{_{kl}}\,\rho_{_{kl}}}\,\Big(\bfm{\mathcal
J}_{_{kl}}[{\vec\rho},\,{\vec S}]+\bfm{\mathcal
R}_{_{kl}}[{\vec\rho},\,{\vec S}]\Big) \ ,\label{cgenf}
\end{equation}
where $\bfm{\mathcal R}_{_{kl}}[{\vec\rho},\,{\vec S}]$ are
arbitrary real functionals related to ${\bfm
G}_{_k}[{\vec\rho},\,{\vec S}]$, introduced in equation
(\ref{cond}), in
\begin{equation}
\sum_{l=1}^{p_{_k}}\bfm{\mathcal R}_{_{kl}}[{\vec\rho},\,{\vec S}]=
{\bfm G}_{_k}[{\vec\rho},\,{\vec S}] \ .\label{conts}
\end{equation}
Consistence of equations (\ref{cgenf}) implies the following
constraints
\begin{equation}
{\bfm\nabla}\times\left[{1\over\rho_{_{kl}}}\Big(\bfm{\mathcal
J}_{_{kl}}[{\vec\rho},\,{\vec S}]+\bfm{\mathcal
R}_{_{kl}}[{\vec\rho},\,{\vec S}]\Big)\right]=0 \ .\label{crot}
\end{equation}
These equations select the potential $U[{\vec\rho},\,{\vec
S}]$ and, through equations (\ref{hermit})-(\ref{antihermitt}), the
nonlinear system where the transformation can be performed. \\
We remark that, according to equations (\ref{cgenf}), equation
(\ref{ctr}) defines a wide class of transformations, one for every
choice of the set of functionals $\bfm{\mathcal
R}_{_{kl}}[{\vec\rho},\,{\vec S}]$. Each transformation changes the
initial system (\ref{c2NLSE}), with the nonlinearity $\widehat
W[{\vec\rho},\,{\vec S}]+i\,\widehat{\mathcal W}[{\vec\rho},\,{\vec
S}]$, in another one with a purely Hermitian nonlinearity $\widehat
W^\prime[{\vec\rho},\,{\vec S}]$.\\ Preliminarily, we observe that,
within the notation (\ref{antientry}) and (\ref{ic}), the matrix
$\widehat{\mathcal W}[\vec\rho,\,\vec S]$ assumes the expression
\begin{equation}
\widehat {\mathcal W}[{\vec\rho},\,{\vec S}]=-{\rm
diag}\left({1\over2\,\rho_{_{kl}}} \Big(I_{_{kl}}[{\vec\rho},\,{\vec
S}]+{\bfm\nabla}\cdot\bfm{\mathcal J}_{_{kl}}[{\vec\rho},\,{\vec
S}]\Big)\right) \ .
\end{equation}
In this way, by performing the gauge transformation, equation
(\ref{c2NLSE}) becomes
\begin{equation}
i\,\frac{\partial\Phi}{\partial t}+\widehat{A}\,\Delta
\Phi+\left(\widehat W_{\rm t}[{\vec\rho},\,{\vec {\mathcal
S}}]+i\,\widehat{\mathcal W}_{\rm t}[{\vec\rho},\,{\vec {\mathcal
S}}]\right)\,\Phi=0 \ ,\label{CNLSE2}
\end{equation}
where
\begin{equation}
\widehat W_{\rm t}[{\vec\rho},\,{\vec {\mathcal S}}]={\rm
diag}\left(w_{_{kl}}- a_{_{kl}}\,\left({\bfm\nabla}\sigma_{_{kl}}
\right)^2+2\,a_{_{kl}}\,{\bfm\nabla}{\mathcal
S}_{_{kl}}\cdot{\bfm\nabla}\sigma_{_{kl}}
+\frac{\partial\sigma_{_{kl}}}{\partial t}\right) \ ,
\end{equation}
and
\begin{equation}
\widehat{\mathcal W}_{\rm t}[{\vec\rho},\,{\vec {\mathcal S}}]={\rm
diag}\left(\frac{1}{2\,\rho_{_{kl}}}{\mathcal
F}_{_{kl}}[{\vec\rho},\,{\vec {\mathcal S}}]\right) \ .
\end{equation}
The functionals ${\mathcal F}_{_{kl}}[{\vec\rho},\,{\vec {\mathcal
S}}]$ are given by
\begin{equation}
{\mathcal F}_{_{kl}}[{\vec\rho},\,{\vec {\mathcal
S}}]=I_{_{kl}}[{\vec\rho},\,{\vec {\mathcal
S}}]-{\bfm\nabla}\cdot\bfm{\mathcal R}_{_{kl}}[{\vec\rho},\,{\vec
{\mathcal S}}] \ ,
\end{equation}
and fulfill the relations
\begin{equation}
\sum_{l=1}^{p_{_k}}{\mathcal F}_{_{kl}}[{\vec\rho},\,{\vec {\mathcal
S}}]=0 \ ,\label{rel}
\end{equation}
as can be verify by using equations (\ref{iic}), (\ref{cond})
and (\ref{conts}).\\
We remark that, as a consequence of this last relation, equation
(\ref{CNLSE2}) admits the following set of continuity equations
\begin{equation}
\frac{\partial\rho_{_k}}{\partial t}+{\bfm\nabla}\cdot{\bfm
j}_{_{\Phi,k}}^{(0)}=0 \ ,
\end{equation}
where
\begin{equation}
{\bfm j}_{_{\Phi,k}}^{(0)}=\sum_{l=1}^{p_k}{\bfm
j}_{_{\Phi,kl}}^{(0)} \ ,\label{jl}
\end{equation}
and
\begin{equation}
{\bfm j}_{_{\Phi,kl}}^{(0)}=2\,a_{_{kl}}
\,\rho_{_{kl}}\,{\bfm\nabla}{\mathcal S}_{_{kl}} \ ,\label{cor1}
\end{equation}
i.e., the nonlinear currents ${\bfm j}_{_{\Psi,k}}$ are transformed
in ${\bfm j}_{_{\Psi,k}}\rightarrow
{\bfm j}_{_{\Phi,k}}^{(0)}$ which have the standard bilinear form.\\
On the other hand, the system (\ref{CNLSE2}), with the nonlinearity
$\widehat W_{\rm t}[{\vec\rho},\,{\vec{\mathcal
S}}]+i\,\widehat{\mathcal W}_{\rm t}[{\vec\rho},\,{\vec{\mathcal
S}}]$, can be rewritten in
\begin{equation}
i\,\frac{\partial\Phi}{\partial t}+\widehat{A}\, \Delta\Phi+\widehat
W^\prime[{\vec\rho},\,{\vec{\mathcal S}}]\,\Phi \ ,\label{CNLSE3}
\end{equation}
with a purely Hermitian nonlinearity $\widehat W^\prime=(\widehat
W^\prime)^\dag$ given in the following block-form
\begin{equation}
\widehat W^\prime[{\vec\rho},\,{\vec{\mathcal S}}]={\rm
diag}\left(\widehat W_{_k}^\prime[{\vec\rho},\,{\vec{\mathcal
S}}]\right) \ .\label{rhermit}
\end{equation}
The $p_{_k}\times p_{_k}$ matrices $\widehat
W^\prime_{_k}[{\vec\rho},\,{\vec{\mathcal S}}]=\widehat
D_{_k}[{\vec\rho},\,{\vec{\mathcal S}}]+\widehat
C_{_k}[{\vec\rho},\,{\vec{\mathcal S}}]$ have a diagonal part
\begin{eqnarray}
\nonumber
\widehat D_{_k}[{\vec\rho},\,{\vec {\mathcal S}}]&=&{\rm
diag}\,\left(w_{_{kl}}-
a_{_{kl}}\,\left({\bfm\nabla}\sigma_{_{kl}}\right)^2+2\,a_{_{kl}}\,
{\bfm\nabla}{\mathcal S}_{_{kl}}\cdot{\bfm\nabla}\sigma_{_{kl}}
+\frac{\partial\sigma_{_{kl}}}{\partial t}\right) \ ,\\\label{d}
\end{eqnarray}
with purely real entries, and an off-diagonal part
\begin{equation}
\left(\widehat C_{_k}\right)_{_{lm}}\!\!\![{\vec\rho},\,{\vec
{\mathcal S}}]=i\,\frac{{\mathcal F}_{_{kl}}-{\mathcal
F}_{_{km}}}{2\, p_{_k}\,\sqrt{\rho_{_{kl}}\, \rho_{_{km}}}}\,e^{i\,
\left(S_{_{kl}}-S_{_{km}}\right)} \ ,\label{c}
\end{equation}
which result to be Hermitian matrices $\widehat C_{_k}=\widehat
C_{_k}^\dag$.\\ We observe that because the Lagrangian
(\ref{qlagrangean}) is U(1)-invariant, the arbitrary integration
constants, deriving from the definition (\ref{cgenf}), do not
produce any effect and can be posed equal to zero. Moreover, the
last term in equation (\ref{d}) can be solved using equations
(\ref{ms})-(\ref{mr}) reducing the nonlinearity $\widehat
W^\prime[\vec\rho,\,\vec S]$ in a quantity containing only space
derivatives.\\
The extension of the method to the case of non canonical coupled
systems is almost immediate and can be performed following the same
steps described at the end of section 3.2.\\
For a non canonical system the matrix $\widehat W[{\vec\rho},\,{\vec
S}]$ can assume any arbitrary expression whereas the form of the
matrix $\widehat{\mathcal W}[{\vec\rho},\,{\vec S}]$ is constrained
by the existence of the set of the continuity equations for the
fields $\rho_{_k}$. Without loss of generality we can pose
\begin{equation}
{\scriptstyle{\mathcal W}}_{_{kl}} [{\vec\rho},\,{\vec
S}]=-{1\over2\,\rho_{_{kl}}}\left(I_{_{kl}}[{\vec\rho},\,{\vec
S}]+{\bfm\nabla}\cdot\bfm{\mathcal J}_{_{kl}}[{\vec\rho},\,{\vec
S}]\right) \ ,\label{wne}
\end{equation}
where now the functionals $\bfm{\mathcal
J}_{_{kl}}[{\vec\rho},\,{\vec S}]$ and $I_{_{kl}}[{\vec\rho},\,{\vec
S}]$ are no more related to the nonlinear potential
$U[{\vec\rho},\,{\vec S}]$ through equations (\ref{antientry}) and
(\ref{ic}). The continuity equations (\ref{qcon}) require that the
functionals $I_{_{kl}}[{\vec\rho},\,{\vec S}]$ still fulfill the
constraints (\ref{cond}) for an arbitrary set of functionals ${\bfm
G}_{_k}[{\vec\rho},\,{\vec S}]$. The total currents ${\bfm
j}_{_{\Psi,k}}$ are given in equation (\ref{ccurrent}) but now the
functionals $I_{_{kl}}[{\vec\rho},\,{\vec S}]$ and $\bfm{\mathcal
J}_{_{kl}}[{\vec\rho},\,{\vec S}]$ are related to the matrix
$\widehat{\mathcal W}[{\vec\rho},\,{\vec S}]$ only through equation
(\ref{wne}). Finally, we introduce the transformation (\ref{ctr})
with generators (\ref{cgenf}) which eliminates the anti-Hermitian
matrix $\widehat{\mathcal W}[{\vec\rho},\,{\vec S}]$ of the
nonlinearity and transforms the system of CNLSEs in the form given
in equation (\ref{CNLSE3}) with only an Hermitian matrix $\widehat
W^\prime[{\vec\rho},\,{\vec S}]$ given still through equations
(\ref{rhermit})-(\ref{c}). \\
Let us now briefly study separately two particular relevant
cases:\\

{\em a) CNLSEs conserving the number of each species of particles}\\

We assume $p=q$ with $p_{_k}=1$ and replace the double index $kl\to
k$. From equation (\ref{c1NLSE}) we obtain the following evolution
equation for the quantities $\rho_{_k}$
\begin{equation}
\frac{\partial\rho_{_k}}{\partial t}+{\bfm\nabla}\cdot{\bfm
j}_{_{\psi,k}}+I_{_k}=0  \ ,
\end{equation}
where the currents ${\bfm j}_{_{\psi,k}}$ are given by
\begin{equation}
\Big({\bfm
j}_{_{\psi,k}}\Big)_i=2\,a_{_k}\,\rho_{_k}\,\partial_iS_{_k}
+\frac{\delta}{\delta(\partial_i\,S_{_k})}\int\limits_{\mathcal R}
U[\vec\rho,\,\vec S]\,d^nx\,dt \ ,
\end{equation}
while the quantities $I_{_k}[{\vec\rho},\,{\vec {\mathcal S}}]$
assume the expression
\begin{equation}
I_{_k}=-\frac{\partial}{\partial S_{_k}}\int\limits_{\mathcal
R}U[{\vec\rho},\,\vec S]\,d^nx\,dt \ .
\end{equation}
Conservation of the single densities $\rho_{_k}$ implies that all
the quantities $I_{_k}[{\vec\rho},\,{\vec {\mathcal S}}]$ must
vanish. This requires that the potential $U[{\vec\rho},\,\vec S]$
depends on the phases $S_{_k}$ only through their spatial
derivatives as it follows also from arguments based on the
invariance of Lagrangian under a global unitary
transformation.\\
Remark that in this case the matrix $\widehat{\mathcal
W}[\vec\rho,\,\vec S]$ assumes the simple expression
\begin{equation}
\widehat{\mathcal W}[\vec\rho,\,\vec S]={\rm
diag}\left(-{1\over2\,\rho_{_k}}{\bfm\nabla}\cdot\bfm{\mathcal
J}_{_k}[\vec\rho,\,\vec S]\right) \ ,\label{antihermit1}
\end{equation}
where the functionals $\bfm{\mathcal J}_{_k}[\vec\rho,\,\vec S]$ are
defined in equations (\ref{jjc}) and (\ref{antientry}), after posing
$p_{_k}=1$. All the quantities ${\bfm G}_{_k}[\vec\rho,\,\vec S]$
introduced in equation (\ref{cond}) reduce to constant vectors
which, without lost of generality, can be posed equal to zero. This
implies that all the functionals ${\mathcal
F}_{_k}[{\vec\rho},\,{\vec {\mathcal S}}]$ vanish and after the
transformation, the matrix $\widehat W^\prime[{\vec\rho},\,{\vec
{\mathcal S}}]$ is reduced in a diagonal form given by
\begin{equation}
\widehat W^\prime[{\vec\rho},\,{\vec {\mathcal S}}]={\rm
diag}\left[w_{_k}- a_{_k}\,\left({\bfm\nabla}\sigma_{_k}\right)^2
+2\,a_{_k}\,{\bfm\nabla}S_{_k}\cdot{\bfm\nabla}\sigma_{_k}
+\frac{\partial\sigma_{_k}}{\partial t}\right] \ ,\label{rhermit1}
\end{equation}
which contains now only a purely real nonlinearity since the
off-diagonal part $\widehat C[{\vec\rho},\,{\vec {\mathcal S}}]$
vanishes. Remark that, in this case, the gauge transformation is
univocally defined because the generators are given by
${\bfm\nabla}\sigma_{_k}[\vec\rho,\,\vec S]={\bfm{\mathcal J}}_{_k}[\vec\rho,\,\vec S]/2\,a_{_k}\,\rho_{_k}$.\\

{\em b) CNLSEs conserving the total number of particles}\\

We pose $q=1$ and replace the double index $kl\to l$. From equation
(\ref{c1NLSE}) we obtain the following evolution equation for the
density $\rho_{_{\rm tot}}$
\begin{equation}
\frac{\partial\rho_{_{\rm tot}}}{\partial t}+{\bfm\nabla}\cdot{\bfm
j}+I_{_{\rm tot}}[\vec\rho,\,\vec S]=0 \ ,\label{cc1}
\end{equation}
where
\begin{equation}
\rho_{_{\rm tot}}=\sum_{l=1}^p\rho_{_l} \ ,
\end{equation}
is the total density of particles and the current $\bfm j$ is given
by
\begin{equation}
\Big({\bfm
j}\Big)_i=\sum_{l=1}^p\left(2\,a_{_l}\,\rho_{_l}\,\partial_iS_{_l}
+\frac{\delta}{\delta(\partial_i\,S_{_l})}\int\limits_{\mathcal R}
U[\vec\rho,\,\vec S]\,d^nx\,dt\right) \ .
\end{equation}
Conservation of $\rho_{_{\rm tot}}$ require that $I_{_{\rm
tot}}[\vec\rho,\,\vec S]$, defined by
\begin{equation}
I_{_{\rm tot}}[\vec\rho,\,\vec
S]=-\sum_{l=1}^p\frac{\partial}{\partial
S_{_l}}\int\limits_{\mathcal R}U[{\vec\rho},\,\vec S]\,d^nx\,dt \ ,
\end{equation}
and can be expressed in
\begin{equation}
I_{_{\rm tot}}[{\vec\rho},\,{\vec S}]={\bfm\nabla}\cdot{\bfm
G}[{\vec\rho},\,{\vec S}] \ ,
\end{equation}
so that equation (\ref{cc1}) becomes a continuity equation
\begin{equation}
\frac{\partial\rho_{_{\rm tot}}}{\partial t}+{\bfm\nabla}\cdot{\bfm
j}_{_{\rm tot}}=0 \ ,
\end{equation}
where the total current ${\bfm j}_{_{\rm tot}}$ is given in
\begin{equation}
{\bfm j}_{_{\rm tot}}={\bfm j}+{\bfm G}[{\vec\rho},\,{\vec S}] \ .
\end{equation}
By performing the transformation (\ref{ctr}) with generator
(\ref{cgenf}), where
\begin{equation}
\sum_{l=1}^p\bfm{\mathcal R}_{_l}[{\vec\rho},\,{\vec S}]={\bfm
G}[{\vec\rho},\,{\vec S}] \ ,
\end{equation}
we obtain the new system of CNLSEs (\ref{CNLSE3}) with an Hermitian
nonlinearity $\widehat{W}^\prime[\vec\rho,\,\vec S]=\widehat
D[\vec\rho,\,\vec S]+\widehat C[\vec\rho,\,\vec S]$. The diagonal
part
\begin{equation}
\widehat D[\vec\rho,\,\vec S]={\rm diag}\,\left[w_{_l}-
a_{_l}\,\left({\bfm\nabla}\sigma_{_l}\right)^2+2\,a_{_l}\,
{\bfm\nabla}{\mathcal S}_{_l}\cdot{\bfm\nabla}\sigma_{_l}
+\frac{\partial\sigma_{_l}}{\partial t}\right] \ ,
\end{equation}
contains purely real entries whilst the off-diagonal part
\begin{equation}
\widehat C_{_{lm}}=i\,\frac{{\mathcal F}_{_l}-{\mathcal
F}_{_m}}{2\,\sqrt{\rho_{_l}\, \rho_{_m}}}\,e^{i\,
\left(S_{_l}-S_{_m}\right)} \ ,
\end{equation}
results to be Hermitian.


\sect{Nonlinear Schr\"odinger equation coupled with gauge fields}

In this section we generalize the nonlinear transformation to NLSEs
coupled with Abelian gauge fields whose dynamic is described by
means of the standard Maxwell term with the inclusion of the additional
Chern-Simons term.

\subsection{The canonical model}

We consider a class of NLSEs describing, in the mean field
approximation, a system of interacting charged particles. The model
is furnished by the following Lagrangian density
\begin{equation}
{\mathcal L}[\psi^\ast,\,\psi,\,A_\mu]={\mathcal L}_{\rm
m}[\psi^\ast,\,\psi,\,A_\mu]+{\mathcal L}_{\rm g}[A_\mu] \
,\label{lagrangianag}
\end{equation}
where the Lagrangian of the matter field ${\mathcal L}_{\rm m}$ is
given by
\begin{eqnarray}
{\mathcal L}_{\rm m}[\psi^\ast,\,\psi,\,A_\mu]
=\frac{i}{2}\,\Big[\psi^\ast\,D_t\psi- \psi\,(D_t\psi)^\ast\Big]-
|{\bfm D}\psi|^2-U[\rho,\,S,\,{\bfm A}] \ ,\label{lagrangecng}
\end{eqnarray}
with $D_\mu=(\partial_\mu+i\,A_\mu)$ the covariant derivative and
$U[\rho,\,S,\,{\bfm A}]$ is the nonlinear potential in the
hydrodynamic representation depending on the abelian gauge field  $A_\mu\equiv(A_0,\,-{\bfm A})$ only through its spatial
components. The Lagrangian of the gauge field ${\mathcal L}_{\rm g}$
assumes the expression
\begin{equation}
{\mathcal L}_{\rm
g}[A_{\mu}]=-\frac{\gamma}{4}\,F_{\mu\nu}\,F^{\mu\nu}+\frac{g}{2}
\,\varepsilon^{\tau\mu\nu}\,A_\tau\,F_{\mu\nu} \ , \label{MCS}
\end{equation}
where $F_{\mu\nu}=\partial_\mu A_\nu-\partial_\nu A_\mu$ is the
electromagnetic tensor, with $\partial_\mu\equiv(\partial/\partial
t,\,{\bfm\nabla})$.\\
In the following, Greek indices take the value $0,\,\ldots,\,n$, the
Latin indices assume the value $1,\,\ldots,\,n$ and denote the
spatial coordinates. Indices are lowered and uppered depending on
the metric tensor $\eta_{\mu\nu}\equiv{\rm
diag}(1,\,-1,\,\ldots,\,-1)$. The Levi-Civita tensor
$\varepsilon^{\tau\mu\nu}$, fully antisymmetric, is defined as
$\varepsilon^{012}=1$. The parameters $\gamma$ and $g$ weight the
contribute of the Maxwell interaction and the Chern-Simons
interaction. We recall that the Chern-Simons term gives contribution
only when the dynamic of the system is constrained in a manifold
with an even number of space dimensions (like in the plane) whilst
in an odd number of space dimensions it reduces to a total
derivative which does not give contribute to the motion equation.\\
Starting from the action of the system
\begin{equation}
{\mathcal A}=\int\limits_{\mathcal R}{\mathcal
L}[\psi^\ast,\,\psi,\,A_\mu]\,d^nx\,dt \ ,
\end{equation}
the evolution equations for the fields $\psi,\,\psi^\ast$ and
$A_{_\mu}$ are obtained by posing $\delta{\mathcal A}=0$ where the
variation is performed with respect
to the 3-vector ${\bfm\Omega}\equiv(\psi,\,\psi^\ast,\,A_\mu)$.\\
The motion equation for the gauge field assumes the expression
\begin{equation}
\gamma\,\partial_\mu F^{\mu\nu}+g\,\varepsilon^{\nu\tau\mu}
\,F_{\tau\mu}=j^\nu_{_{A\psi}} \ ,  \label{gaugefield}
\end{equation}
where the covariant current $j^\nu_{_{A\psi}}\equiv(\rho,\,{\bfm
j}_{_{A\psi}})$ has spatial components
\begin{eqnarray}
\Big({\bfm
j}_{_{A\psi}}\Big)_i=2\,\rho\,\Big(\partial_iS+A_i\Big)+\frac{\delta}{\delta
A_i}\int\limits_{\mathcal R} U[\rho,\,S,\,{\bfm A}]\,d^nx\,dt \
.\label{ccg}
\end{eqnarray}
By observing that $F^{\mu\nu}=-F^{\nu\mu}$, from equation
(\ref{gaugefield}) we immediately obtain the continuity equation for
the field $\rho$
\begin{equation}
\frac{\partial\rho}{\partial t}+{\bfm\nabla}\cdot{\bfm
j}_{_{A\psi}}=0 \ ,\label{continuity}
\end{equation}
which assures the conservation of the total charge of the system.\\
On the other hand, the evolution equation for the matter field, as
it follows from the Lagrangian density (\ref{lagrangianag}), is
given by
\begin{equation}
i\,D_t\psi+{\bfm D}^2\psi+\Big(W[\rho,\,S,\,{\bfm A}]+i\,{\mathcal
W}[\rho,\,S,\,{\bfm A}]\Big)\,\psi=0 \ ,\label{schroedingerg2}
\end{equation}
where the real and imaginary parts of the nonlinearity are given,
respectively,  by
\begin{equation}
W[\rho,\,S,\,{\bfm
A}]=-\frac{\delta}{\delta\rho}\,\int\limits_{\mathcal R}
U[\rho,\,S,\,{\bfm A}]\,d^nx\,dt \ ,
\end{equation}
and
\begin{equation}
{\mathcal W}[\rho,\,S,\,{\bfm
A}]=-\frac{1}{2\,\rho}\,\frac{\delta}{\delta
S}\,\int\limits_{\mathcal R} U[\rho,\,S,\,{\bfm A}]\,d^nx\,dt \
.\label{wa}
\end{equation}
For consistence, equation (\ref{schroedingerg2}) must admit
the same continuity equation (\ref{continuity}).\\
Following standard arguments, by multiply equation
(\ref{schroedingerg2}) by $\psi^\ast$  and taking its imaginary
part, we obtain
\begin{equation}
\frac{\partial\rho}{\partial
t}+{\bfm\nabla}\cdot\Big[2\,\rho\left({\bfm\nabla}S-{\bfm
A}\right)\Big] -\frac{\delta}{\delta S}\int\limits_{\mathcal R}
U[\rho,\,S,\,{\bfm A}]\,d^nx\,dt=0 \ ,
\end{equation}
which can be written in
\begin{equation}
\frac{\partial\rho}{\partial t}+{\bfm\nabla}\cdot{\bfm
j}_{_{A\psi}}=\frac{\partial}{\partial S}\int\limits_{\mathcal R}
U[\rho,\,S,\,{\bfm A}]\,d^nx\,dt \ ,\label{eee}
\end{equation}
where the charged current ${\bfm j}_{_{A\psi}}$ now becomes
\begin{equation}
\Big({\bfm
j}_{_{A\psi}}\Big)_i=2\,\rho\left(\partial_iS+A_i\right)+\frac{\delta}{\delta(\partial_iS)}
\,\int\limits_{\mathcal R} U[\rho,\,S,\,{\bfm A}]\,d^nx\,dt \
.\label{currentgg}
\end{equation}
By comparing this expression with equation (\ref{ccg}) it follows
that:\\
1) the nonlinear potential $U[\rho,\,S,\,{\bfm A}]$ must depend on
the field $S$ only through its spatial derivatives so that the right
hand side of equation (\ref{eee}) vanishes becoming, in this way, a
continuity equation for the field $\rho$.\\
2) the fields ${\bfm \nabla}S$ and $\bfm A$ must be present in the
nonlinear potential through the combination ${\bfm\nabla}S-{\bfm
A}$. In other words the Lagrangian of the matter field can be
obtained consistently from the Lagrangian of the scalar field
(\ref{lagrangean}) by replacing in it the standard derivatives with the
covariant ones $\partial_\mu\rightarrow D_\mu=\partial_\mu+i\,A_\mu$
(minimal coupling prescription).\\
Since, as a required, $U[\rho,\,S,\,{\bfm A}]$ depends only through
the quantity ${\bfm\nabla}S-{\bfm A}$ and its highest spatial
derivatives, equation (\ref{wa}) can be written in
\begin{equation}
{\mathcal W}[\rho,\,S,\,{\bfm
A}]={1\over2\,\rho}\,{\bfm\nabla}\cdot{\bfm{\mathcal
J}_{\!\!A}}[\rho,\,S,\,{\bfm A}] \ ,
\end{equation}
where the vector $\bfm{\mathcal J}_{\!\!A}[\rho,\,S,\,{\bfm A}]$ is
defined in
\begin{equation}
\Big({\mathcal J}_{\!A}\Big)_i[\rho,\,S,\,{\bfm
A}]=\frac{\delta}{\delta(\partial_iS)}\int\limits_{\mathcal R}
U[\rho,\,S,\,{\bfm A}]\,d^nx\,dt \ ,
\end{equation}
and the charged current (\ref{currentgg}) assumes the expresion
\begin{equation}
{\bfm j}_{_{A\psi}}=2\,\rho\left({\bfm\nabla}S-{\bfm
A}\right)+\bfm{\mathcal J}_{\!\!A}[\rho,\,S,\,{\bfm A}] \ .
\end{equation}

\subsection{Gauge transformation}
Firstly, we recall that the system described by the Lagrangian
(\ref{lagrangianag}) is invariant over a local U(1) transformation
(gauge transformation of second kind), accomplished both on the
fields $\psi$ and $A_\mu$, by means of
\begin{eqnarray}
\nonumber &&A_{\mu}({\bfm x},\,t)\rightarrow A_{\mu}({\bfm x},\,t)
-\partial_{\mu}\omega({\bfm x},\,t) \ ,\\
&&\label{gauge}\\
\nonumber &&\psi({\bfm x},\,t)\rightarrow \exp\Big(i\,\omega({\bfm
x},\,t)\Big)\psi({\bfm x},\,t) \ ,
\end{eqnarray}
where $\omega({\bfm x},\,t)$ is a well-behaved function in the sense
of $\epsilon^{\mu\nu}
\partial_\mu\partial_\nu\omega=0$, with $\epsilon^{\mu\nu}$ the anti-symmetric
 tensor $\epsilon^{\mu\nu}=-\epsilon^{
\nu\mu}$. Remark that, under the transformation (\ref{gauge}),
Lagrangian (\ref{MCS}) changes according to
\begin{equation}
{\mathcal L}_{\rm g}\rightarrow{\mathcal L}_{\rm
g}+\frac{g}{2}\,\epsilon^{\mu\nu\tau}\,\partial_\mu
\left(\omega\,F_{\nu\tau}\right)
 \ ,
\end{equation}
with an extra surface term which does not change the motion of
equations for the fields $\psi$ and $A_\mu$.\\
Let us now introduce the gauge transformation of third kind as a
unitary nonlinear transformation performed only on the field $\psi$
\begin{equation}
\psi({\bfm x},\,t)\rightarrow\phi({\bfm x},\,t)={\mathcal
U}[\rho,\,S,\,{\bfm A}]\,\psi({\bfm x},\,t) \ ,\label{trasf1g}
\end{equation}
which allows to eliminate the imaginary part
$\mathcal W[\rho,\,S,\,{\bfm A}]$ of the nonlinearity in the evolution equation
(\ref{schroedingerg2}) and reduces the charged current to the
standard bilinear form
\begin{equation}
{\bfm j}_{_{A\psi}}[\rho,\,S,\,{\bfm A}]\rightarrow{\bfm
j}^{(0)}_{_{A\phi}} [\rho,\,{\mathcal S},\,{\bfm A}]=2\,\rho\,({\bfm
\nabla}{\mathcal S}-{\bfm A}) \ .\label{currentg}
\end{equation}
The unitary functional ${\mathcal U}[\rho,\,S,\,{\bfm A}]$ is
given by
\begin{equation}
{\mathcal U}[\rho,\,S,\,{\bfm
A}]=\exp\Big(i\,\sigma\left[\rho,\,S,\,{\bfm A}\right]\Big)
 \ ,\label{trasf2g}
\end{equation}
where the real generator of the transformation
$\sigma\left[\rho,\,S,\,{\bfm A}\right]$ defined according to
\begin{equation}
{\bfm\nabla}\sigma\left[\rho,\,S,\,{\bfm
A}\right]=\frac{1}{2\,\rho}\,\bfm{\mathcal
J}_{\!\!A}[\rho,\,S,\,{\bfm A}] \ ,\label{sig}
\end{equation}
are constrained by
\begin{equation}
{\bfm\nabla}\times\left(\frac{\bfm{\mathcal
J}_{\!\!A}[\rho,\,S,\,{\bfm A}]}{\rho}\right)=0 \
.\label{condition1}
\end{equation}
By performing the transformation (\ref{trasf1g}), from equation
(\ref{schroedingerg2}) we obtain the following NLSE for the charged
field $\phi$
\begin{equation}
i\,D_t\phi+{\bfm D}^2\phi+\widetilde{W}[\rho,\,{\mathcal S},\,{\bfm
A}]\,\phi=0 \ ,\label{schroedinger2g}
\end{equation}
where the real nonlinearity $\widetilde W[\rho,\,S,\,{\bfm A}]$
assumes the expression
\begin{equation}
\widetilde{W}[\rho,\,{\mathcal S},\,{\bfm
A}]=W-({\bfm\nabla}\sigma)^2+2\,(\bfm{\nabla}{\mathcal S}-{\bfm A})
\cdot{\bfm\nabla}\sigma+\frac{\partial\sigma}{\partial t} \ ,
\end{equation}
with $W\equiv W[\rho,\,S[\rho,\,{\mathcal S},\,{\bfm A}],\,{\bfm
A}]$ and $\sigma\equiv \sigma[\rho,\,S[\rho,\,{\mathcal S},\,{\bfm
A}],\,{\bfm A}]$. The new phase $\mathcal S$ of the field $\phi$ is
related to the old phase $S$ of the field $\psi$ through the
relation
\begin{equation}
{\mathcal S}=S+\sigma[\rho,\,S,\,{\bfm A}] \ ,\label{ns}
\end{equation}
and because the nonlinearity in equation (\ref{schroedinger2g}) is a
purely real quantity the continuity equation for the field $\rho$
becomes
\begin{equation}
\frac{\partial\rho}{\partial t}+{\bfm\nabla}\cdot{\bfm
j}^{(0)}_{_{A\phi}}=0 \ ,
\end{equation}
with the transformed charged current ${\bfm j}^{(0)}_{_{A\phi}}$ given in equation (\ref{currentg}).\\
Since the nonlinear transformation has been accomplished only on the
matter field, the evolution equation for the gauge field retains
formally the same expression given in equation (\ref{gaugefield})
\begin{equation}
\gamma\,\partial_\mu\,F^{\mu\nu}+g\,\varepsilon^{\nu\tau\mu}
\,F_{\tau\mu}=j^\nu_{_{A\phi}} \ ,
\end{equation}
but with the transformed charged source
$j^\nu_{_{A\phi}}\equiv(\rho,\,{\bfm j}^{(0)}_{_{A\phi}})$. \\On the other
hand, the presence of the gauge field enable us to
introduce a transformation on it, leaving the matter field unchanged.\\
In fact, let us introduce the following transformation
\begin{equation}
{\bfm A}({\bfm x},\,t)\rightarrow{\bfm\chi}({\bfm x},\,t)={\bfm
A}({\bfm x},\,t)-{\bfm\nabla}\sigma[\rho,\,S,\,{\bfm A}] \ ,
\label{trasfgauge1}
\end{equation}
where $\sigma[\rho,\,S,\,{\bfm A}]$ is still defined through
equation
(\ref{sig}).\\
Accounting for $F_{\mu\nu}=-F_{\nu\mu}$, it follows that
\begin{equation}
F_{\mu\nu}\equiv\partial_\mu A_\nu-\partial_\nu A_\mu=
\partial_\mu\chi_\nu-\partial_\nu\chi_\mu \
,\label{condition2}
\end{equation}
whenever $\sigma[\rho,\,S,\,{\bfm A}]$ is a well behaved function
fulfilling the relation
\begin{equation}
\epsilon^{\mu\nu}\,\partial_\mu\partial_\nu\sigma[\rho,\,S,\,{\bfm
A}]=0 \ .\label{trasfgauge2}
\end{equation}
This implies that, for $\mu$ and $\nu$ spatial indices, equation
(\ref{condition2}) is trivially satisfied as consequence of
condition (\ref{condition1}), differently, for $\mu$ or $\nu$ equal
to zero, equation (\ref{condition2}) implies the following
transformation for the component $A_0({\bfm x},\,t)$ of the gauge
field
\begin{equation}
A_0({\bfm x},\,t)\rightarrow\chi_0({\bfm x},\,t)=A_0({\bfm
x},\,t)+\frac{\partial}{\partial t}\,\sigma[\rho,\,S,\,{\bfm A}] \ .
\label{trasfgauge3}
\end{equation}
By performing the transformation (\ref{trasfgauge1}) and
(\ref{trasfgauge3}) in equation (\ref{gaugefield}) we obtain
\begin{equation}
\gamma\,\partial_\mu F^{\mu\nu}+g\,\varepsilon^{\nu\tau\mu}
\,F_{\tau\mu}=\tilde j^\nu_{_{A\phi}} \ ,
\end{equation}
where the new covariant current $\tilde
j^\nu_{_{A\phi}}\equiv(\rho,\,\tilde {\bfm j}^{(0)}_{_{A\phi}})$ with
\begin{equation}
\tilde {\bfm j}^{(0)}_{_{A\phi}}=2\,\rho\,({\bfm\nabla}S-{\bfm\chi}) \ ,
\end{equation}
fulfills the continuity equation
\begin{equation}
\frac{\partial\rho}{\partial t}+{\bfm\nabla}\cdot\tilde {\bfm
j}^{(0)}_{_{A\phi}}=0 \ .
\end{equation}
Differently, from equation (\ref{schroedingerg2}) it follows
\begin{equation}
i\,{\overline D}_t\psi+\overline{\bfm
D}^{\,2}\psi+\overline{W}[\rho,\,S,\,{\bfm \chi}]\,\psi=0 \ ,
\end{equation}
which has the same form of equation (\ref{schroedinger2g}) but now
the covariant derivative is defined in
$\overline{D}_\mu=\partial_\mu+i\,\chi_\mu$, while the real
nonlinearity becomes
\begin{equation}
\overline{W}[\rho,\,S,\,{\bfm\chi}]=W-({\bfm\nabla}\sigma)^2+2\,({\bfm
\nabla}S-{\bfm\chi})\cdot{\bfm\nabla}\sigma+\frac{\partial\sigma}{\partial
t} \ ,
\end{equation}
with $W\equiv W[\rho,\,S,\,{\bfm A}[\rho,\,S,\,{\bfm\chi}]]$ and
$\sigma\equiv \sigma[\rho,\,S,\,{\bfm A}[\rho,\,S,\,{\bfm\chi}]]$.\\
In conclusion, it is worthy to observe that if we introduce the
nonlinear transformation both on the matter field and the gauge
filed, by following the prescription given in equation
(\ref{gauge}), the evolution equations (\ref{gaugefield}) and
(\ref{schroedingerg2}) are not changed in form because the
variations due to the matter field are balanced by the variations
due to the gauge field. Thus, in this case transformation
(\ref{trasf1g}) behaviors exactly like a gauge transformation of
second kind.


\sect{Applications} To show the applicability of the nonlinear
transformation introduced in this paper, we consider some examples
for the three cases: scalar NLSEs, coupled NLSEs and gauged NLSEs.
Some of the examples here discussed are already known in literature.
We show that the nonlinear transformations introduced by different
Authors can be obtained, in a unified way, with the method presented
in this work.

\subsection{Scalar NLSEs}

Let us consider, as a first example, the following 1-dimensional
NLSE
\begin{eqnarray}
\nonumber i\,\frac{\partial\psi}{\partial
t}&+&\frac{\partial^2\psi}{\partial x^2}
+a_{_1}\,|\psi|^2\,\psi+a_{_2}\,|\psi|^4\,\psi\\
&+&i\,a_{_3}\,|\psi|^2\,\frac{\partial\psi}{\partial x}\,\psi
+i\,a_{_4}\,\frac{\partial\psi^\ast}{\partial x}\,\psi^2=0 \
,\label{exe1}
\end{eqnarray}
where $a_{_1}$, $a_{_2}$, $a_{_3}$ and $a_{_4}$ are real constants.
After introducing the hydrodynamic fields $\rho$ and $S$ we can
write the real and imaginary part of the nonlinearity in
\begin{equation}
W[\rho,\,S]=b_{_1}\,\rho+b_{_2}\,\rho^2+b_{_3}\,\rho\,\frac{\partial
S}{\partial x} \ ,
\end{equation}
and
\begin{equation}
{\mathcal W}[\rho]=b_{_4}\,\frac{\partial\rho}{\partial x} \ ,
\end{equation}
where $b_{_1}=a_{_1}$, $b_{_2}=a_{_2}$, $b_{_3}=a_{_4}-a_{_3}$ and
$b_{_4}=(a_{_3}+a_{_4})/2$.\\
The canonical subclass of equation (\ref{exe1}) is given by posing
$b_{_3}=-2\,b_{_4}$ and admits the following potential
\begin{equation}
U[\rho,\,S]=-\left({b_{_1}\over2}\,\rho^2+{b_{_2}\over3}\,\rho^3+
{b_{_3}\over2}\,\rho^2\,\frac{\partial S}{\partial x} \right) \ .
\end{equation}
Equation (\ref{exe1}) conserves the density $\rho$ and the
corresponding particles current is given by
\begin{equation}
j_{_\psi}=2\,\rho\,\frac{\partial S}{\partial x}+b_{_4}\,\rho^2 \ .
\end{equation}
After performing the transformation (\ref{transf1}) with
\begin{equation}
\sigma[\rho]={b_{_4}\over2}\int\limits^x\rho\,dx^\prime \ ,
\end{equation}
equation (\ref{exe1}) is changed in
\begin{equation}
i\,\frac{\partial\phi}{\partial t}+\frac{\partial^2\phi}{\partial
x^2} +\left(b_{_1}\,\rho+\tilde
{b}_{_2}\,\rho^2+b_{_3}\,\rho\,\frac{\partial{\mathcal S}}{\partial
x}\right)\,\phi=0 \ ,\label{exetrasf1}
\end{equation}
where $\tilde{b}_{_2}=b_{_2}-b_{_3}\,b_{_4}/2-b_{_4}^2/4$.\\
Equation (\ref{exe1}) contains, as particular cases, some know
NLSEs. Among them we recall:\\
1) The Chen-Lee-Liu equation \cite{Chen} ($b_{_1}=b_{_2}=0$,
$b_{_3}=-2\,b_{_4}$) which is transformed in the NLSE with real
nonlinearity
\begin{equation}
\widetilde W[\rho,\,{\mathcal S}]=\tilde
b_{_2}\rho^2+b_{_3}\,\rho\,\frac{\partial{\mathcal S}}{\partial x} \
,
\end{equation}
where $\tilde b_{_2}=3\,b_{_3}^2/4$.\\
2) The Jackiw-Aglietti equation \cite{Aglietti,Jackiw} ($b_{_1}=0$,
$b_{_2}=-3\,b_{_4}/4$ and $b_{_3}=-2\,b_{_4}$) which is transformed
in the NLSE with real nonlinearity
\begin{equation}
\widetilde W[\rho,\,{\mathcal
S}]=b_{_3}\,\rho\,\frac{\partial{\mathcal S}}{\partial x} \ .
\end{equation}
3) The Eckaus equation \cite{Calogero6,Calogero1}
($b_{_1}=b_{_3}=0$)
 which is transformed in the NLSE with real
nonlinearity
\begin{equation}
\widetilde W[\rho]=\tilde b_{_2}\rho^2 \ ,
\end{equation}
with $\tilde b_{_2}=b_{_2}-b_{_4}^2/4$. Remark that, when
$b_{_1}\not=0$ we obtain, after transformation, the cubic-quintic
NLSE with real nonlinearity
\begin{equation}
\widetilde W[\rho]=b_{_1}\,\rho+\tilde b_{_2}\rho^2 \ ,
\end{equation}
studied in \cite{Ginsburg}.\\
4) The Kaup-Newell equation \cite{Kaup} ($b_{_1}=b_{_2}=0$ and
$b_{_4}=-3\,b_{_3}/2$) which is transformed in the NLSE with real
nonlinearity
\begin{equation}
\tilde W[\rho,\,{\mathcal S}]=\tilde b_{_2}\rho^2+
b_{_3}\,\rho\,\frac{\partial{\mathcal S}}{\partial x} \ ,
\end{equation}
with $\tilde b_{_2}=3\,b_{_3}^2/16$.\\
As a second example we consider the canonical NLSE introduced in
\cite{Kaniadakis,Kaniadakis1}
\begin{eqnarray}
\nonumber i\,\frac{\partial\psi}{\partial
t}+\frac{\partial^2\psi}{\partial x^2}
&+&\kappa\,\left(\psi^\ast\,\frac{\partial\psi}{\partial x}
-\psi\,\frac{\partial\psi^\ast}{\partial x}\right)\,
\frac{\partial\psi}{\partial x}\\
&+&\frac{\kappa}{2}\, \frac{\partial}{\partial x}\,\left
(\psi^\ast\frac{\partial \psi}{\partial
x}-\psi\,\frac{\partial\psi^\ast}{\partial x}\right) \,\psi=0 \
,\label{exe2}
\end{eqnarray}
where $\kappa$ is a real parameter. The real and imaginary
nonlinearities in the hydrodynamic representation are given by
\begin{equation}
W[\rho,\,S]= -2\,\kappa\,\rho\,\left(\frac{\partial S}{\partial
x}\right)^2 \ ,
\end{equation}
and
\begin{equation}
{\mathcal W}[\rho,\,S]=
\frac{\kappa}{\rho}\,\frac{\partial}{\partial x}
\left(\rho^2\,\frac{\partial S}{\partial x}\right) \ .
\end{equation}
They are obtained from the potential
\begin{equation}
U[\rho,\,S]=\kappa\,\left(\rho\,\frac{\partial S}{\partial
x}\right)^2 \ ,\label{ueip}
\end{equation}
whereas the particles current assumes the expression
\begin{equation}
j_{_\psi}=2\,\rho\,(1+\kappa\,\rho)\,\frac{\partial S}{\partial x} \
.
\end{equation}
By performing the transformation (\ref{transf1}) with generator
\begin{equation}
\sigma[\rho,\,S]=\kappa\int\limits^x\rho\,\frac{\partial S}{\partial
x^\prime}\,dx^\prime \ ,\label{sigma2}
\end{equation}
equation (\ref{exe2}) changes in
\begin{equation}
i\,\frac{\partial\phi}{\partial t}+\frac{\partial^2\phi}{\partial
x^2}-\left[
\frac{2\,\kappa\,\rho}{1+\kappa\,\rho}\,\left(\frac{\partial{\mathcal
S}}{\partial x}\right)^2
-\frac{\kappa}{2}\,\rho\,\frac{\partial^2\log\rho}{\partial
x^2}\right]\,\phi=0 \ .\label{exe21}
\end{equation}
Remark that although equation (\ref{exe2}) can be generalized in any
spatial dimensions \cite{Kaniadakis,Kaniadakis1} condition
(\ref{rot}) is not satisfied in general and the transformation
(\ref{transf1}) can be applied consistently only in
1-dimensional case.\\
Another example is given by the class of the Doebner-Goldin
equations \cite{Doebner4}
\begin{equation}
i\,\frac{\partial\,\psi}{\partial\,t}+\Delta\,\psi+\left(\sum_{i=1}^5
c_{_i}\,R_{_i}[\rho,\,S]+i\,
\frac{D}{2}\,R_{_2}[\rho]\right)\,\psi=0 \ ,\label{exe3}
\end{equation}
where the nonlinear functionals $R_{_i}$ are given by
$R_{_1}={\bfm\nabla}\cdot(\rho{\bfm \nabla}S)/\rho$,
$R_{_2}=\Delta\rho/\rho$, $R_{_3}=({\bfm \nabla}S)^2$, $R_{_4}={\bfm
\nabla}S\cdot{\bfm\nabla}\rho/\rho$ and
$R_{_5}=({\bfm\nabla}\rho/\rho)^2$. The canonical subclass of
equation (\ref{exe3}) is obtained for $c_{_1}=-c_{_4}=D$, $c_{_3}=0$
and $c_{_2}=-2\,c_{_5}$ and it follows from the potential
\begin{equation}
U[\rho,\,S] =D\,{\bfm\nabla}\rho\cdot
{\bfm\nabla}S+c_{_5}\,\frac{({\bfm\nabla}\rho)^2}{\rho} \ .
\end{equation}
The particles current is given by
\begin{equation}
{\bfm j}_{_\psi}=2\,\rho\,{\bfm\nabla}S+D\,{\bfm\nabla}\rho \
,\label{currentexe3}
\end{equation}
and the corresponding continuity equation is the well-known
Fokker-Planck equation
\begin{equation}
\frac{\partial\rho}{\partial t}+{\bfm\nabla}\cdot{\bfm
j}_{_\psi}^{(0)}+D\,\Delta\rho=0 \ ,
\end{equation}
where $D$ is the diffusion coefficient.\\ By performing the
transformation (\ref{transf1}) with generator
\begin{equation}
\sigma[\rho]={D\over2}\,\log\rho \ ,\label{ss}
\end{equation}
equation (\ref{exe3}) transforms in
\begin{equation}
i\,\frac{\partial\phi}{\partial t}+\Delta\phi+\sum_{i=1}^5
\tilde{c}_{_i}\,R_{_i}[\rho,\,{\mathcal S}]\,\phi=0 \ ,
\end{equation}
with coefficients $\tilde{c}_{_1}=c_{_1}-D$,
$\tilde{c}_{_2}=c_{_2}-c_{_1}\,D/2$, $\tilde{c}_{_3}=c_{_3}$,
$\tilde{c}_{_4}=c_{_4}+(c_{_3}-1)\,D$ and
$\tilde{c}_{_5}=c_{_5}-c_{_4}\,D-(c_{_3}-1)\,D^2/4$.\\
It is easy to verify that the generator (\ref{ss}) satisfies
condition (\ref{rot}) and the nonlinear transformation can be
performed in any $n\geq1$ spatial
dimensions.\\
As a final example we consider the following family of NLSE
\begin{equation}
i\,\frac{\partial\psi}{\partial
t}+\Delta\psi+\Big(W(\rho,\,S)+i\,{\mathcal
W}(\rho,\,S)\Big)\,\psi=0 \ ,\label{sent}
\end{equation}
with nonlinearities
\begin{equation}
W(\rho,\,S)=-{D\over2}\,f(\rho)\,{\bfm\nabla}\cdot\left(\frac{{\bfm
j}^{(0)}_{_\psi}}{\rho}\right)+G[\rho] \ , \label{www}
\end{equation}
and
\begin{equation}
{\mathcal
W}(\rho,\,S)=-\frac{D}{2\,\rho}\,{\bfm\nabla}\cdot\Big(f(\rho )
\,{\bfm\nabla}\rho\Big) \ , \label{cwww}
\end{equation}
where
\begin{equation}
f(\rho)=\rho\,\frac{\partial\ln\,\kappa(\rho)}{\partial\rho} \ ,
\end{equation}
and $G[\rho]$ is an arbitrary functional of $\rho$. Equation
(\ref{sent}) can be obtained from the potential
\begin{equation}
U[\rho,\,S]=-D\,f(\rho)\,{\bfm\nabla}\rho\cdot{\bfm\nabla}S+\int\limits^\rho
G[\rho^\prime]\,d\rho^\prime \ ,
\end{equation}
and was recently derived in the canonical quantization framework
from a classical many body systems described by generalized
entropies \cite{Scarfone8}.\\ The particles current is given by
\begin{equation}
{\bfm j}_{_\psi}=2\,\rho\,{\bfm
\nabla}\,S-D\,f(\rho)\,{\bfm\nabla}\rho \ ,\label{ncurrent}
\end{equation}
which is the sum of a linear drift current ${\bfm j}_{\rm
drift}=2\,\rho\,{\bfm \nabla}S$ and a nonlinear diffusion current
${\bfm j}_{\rm diff}=-D\,f(\rho)\,{\bfm \nabla}\rho$ different from
Fick's current ${\bfm j}_{\rm Fick}=-D\,{\bfm\nabla}\,\rho$ which is
recovered by posing $\kappa(\rho)=\alpha\,\rho$, with $\alpha$ a
constant. The diffusive term is related to the entropy of the
classical system through the relation (with the Boltzmann constant
$k_{\rm B}=1$)
\begin{equation}
S(\rho)=-\int\limits_M d^nx\int\limits^\rho
\ln\kappa(\rho^\prime)\,d\rho^\prime \ .\label{entropy}
\end{equation}
By performing the transformation (\ref{transf1}) with generator
\begin{equation}
\sigma[\rho]=\frac{D}{2}\,\ln\kappa(\rho) \ ,
\end{equation}
equation (\ref{sent}) changes in
\begin{equation}
i\,\frac{\partial\phi}{\partial t}+\Delta\phi -{D^2\over2}\,\Bigg[
f_{1}(\rho)\,\Delta\rho+ f_2(\rho)\,
\left({\bfm\nabla}\rho\right)^2\Bigg]\,\phi+G[\rho]\,\phi=0 \ ,
\end{equation}
with
\begin{eqnarray}
&&
f_{1}(\rho)=\rho\,\left(\frac{\partial}{\partial\rho}\,\ln\kappa(\rho)
\right)^2 \ ,\\
&& f_2(\rho)={1\over2}\,\frac{\partial\, f_{1}(\rho)}{\partial\rho}
\ ,
\end{eqnarray}
which contains a purely real nonlinearity depending only on $\rho$.\\
In particular, starting from the entropy $S=-\int_M\rho\,\log\rho\,d^nx$, with $\kappa(\rho)=e\,\rho$,
equation (\ref{sent}) becomes
\begin{equation}
i\,\frac{\partial\psi}{\partial t}+\Delta\psi-{D\over
2}\,{\bfm\nabla}\cdot\left(\frac{{\bfm
j}^{(0)}_{_\psi}}{\rho}\right)\,\psi-i\,\frac{D}{2}\,\frac{\Delta\rho}{\rho}\,\psi=0
\ ,\label{DG}
\end{equation}
which coincides with the canonical sub-family of the Doebner-Goldin
equations described in the previous example. After transformation it
becomes
\begin{equation}
i\,\frac{\partial\,\phi}{\partial\,t}+\Delta\,\phi
-{D^2\over2}\,\left[\frac{\Delta\rho}{\rho}
-{1\over2}\left(\frac{{\bfm\nabla}\rho}{\rho}\right)^2\right]\,\phi=0
\ ,\label{DG1}
\end{equation}
which was studied previously in \cite{Guerra}. Remarkably, this
equation is equivalents to the following linear Schr\"odinger
equation
\begin{equation}
i\,k^{\!\!\!\!\!-}\,\frac{\partial\chi}{\partial
t}+{k^{\!\!\!\!\!-}}^2\,\Delta\chi=0 \ ,\label{lsch}
\end{equation}
with $k^{\!\!\!\!\!-}=\sqrt{1-D^2}$ where the field $\chi$ is
related to hydrodynamic fields $\rho$ and $\mathcal S$ through the
relation
$\chi=\sqrt{\rho}\,\exp(i\,{\mathcal S}/k^{\!\!\!\!\!-})$.\\


\subsection{Coupled NLSEs}

Let us now consider the following 1-dimensional system of CNLSEs
\begin{equation}
i\,\frac{\partial\psi_{_j}}{\partial
t}+a_{_j}\,\frac{\partial^2\psi_{_j}}{\partial x^2}
+\Lambda(\psi_{_i},\,\psi^\ast_{_i})\,\psi_{_j}=0 \ ,\label{example}
\end{equation}
with nonlinearity
\begin{eqnarray}
\nonumber \Lambda(\psi_{_i},\,\psi^\ast_{_i})=-i\sum_{i=1}^p&&\Bigg(\alpha_{_{ij}}\frac{\rho_{_i}}{\rho_{_j}}\,
\psi_{_j}\frac{\partial\psi_{_j}^\ast}{\partial x}
+\beta_{_{ij}}\,\rho_{_i}\,\frac{\partial}{\partial
x}\\
&&+\gamma_{_{ij}}\,\psi_{_i}\frac{\partial\psi_{_i}^\ast}{\partial
x}
+\epsilon_{_{ij}}\,\psi_{_i}^\ast\frac{\partial\psi_{_i}}{\partial
x}+f_{_j}({\vec \rho})\Bigg) \ , \label{nl}
\end{eqnarray}
where $\alpha_{_{ij}}$, $\beta_{_{ij}}$, $\gamma_{_{ij}}$ and
$\epsilon_{_{ij}}$ are real constants and $f_{_j}({\vec\rho})$ are
arbitrary real
functionals depending only on the vector field $\vec \rho$.\\
In the hydrodynamic representation the nonlinearity (\ref{nl}) has
Hermitian and anti-Hermitian part given, respectively, by
\begin{eqnarray}
\widehat W[{\vec\rho},\,{\vec S}]&=&{\rm
diag}\left[\sum_{i=1}^p\rho_{_i}\,\left(b_{_{ij}}\, \frac{\partial
S_{_j}}{\partial x}+c_{_{ij}}\,\frac{\partial
S_{_i}}{\partial x}\right)+f_{_j}(\vec \rho)\right] \ ,\label{ew1}\\
\widehat{\mathcal W}[{\vec\rho}]&=&{\rm diag}\left[\sum_{i=1}^p
\left(d_{_{ij}}\,\frac{\rho_{_i}}{\rho_{_j}}\,
\frac{\partial\rho_{_j}}{\partial
x}+e_{_{ij}}\,\frac{\partial\rho_{_i}}{\partial x}\right)\right] \
,\label{ew2}
\end{eqnarray}
where $b_{_{ij}}=\alpha_{_{ij}}-\beta_{_{ij}}$,
$c_{_{ij}}=\gamma_{_{ij}}-\epsilon_{_{ij}}$,
$d_{_{ij}}=(\alpha_{_{ij}}+\beta_{_{ij}})/2$ and
$e_{_{ij}}=(\gamma_{_{ij}}+\epsilon_{_{ij}})/2$.\\Equation
(\ref{example}) includes some cases already known in literature. For
instance: the vector generalization of the Kaup-Newell equation
\cite{Fordy} ($a_{_j}=1$, $c_{_{ij}}=0$,
$-b_{_{ij}}=2\,d_{_{ij}}=e_{_{ij}}=\beta$ and
$f_{_j}({\vec\rho})=0$); the coupled Chen-Lie-Liu equation (Type I)
\cite{Wadati} ($a_{_j}=1$, $c_{_{ij}}=e_{_{ij}}=0$,
$-b_{_{ij}}=2\,d_{_{ij}}=\beta$, $f_{_j}({\vec\rho})=0$); the
coupled Chen-Lie-Liu equation (Type II)  \cite{Wadati} ($a_{_j}=1$,
$b_{_{ij}}=d_{_{ij}}=0$, $c_{_{ij}}=-2\,e_{_{ij}}=\beta$,
$f_{_j}({\vec\rho})=0$); the hybrid CNLSE \cite{Hisakado2,Hisakado1}
($a_{_j}=1$, $c_{_{ij}}=0$,
$-b_{_{ij}}=2\,d_{_{ij}}=e_{_{ij}}=\beta$ and $f_{_j}({\vec
\rho})=\beta\sum_k\rho_{_k}$); the vectorial Eckhaus equation
\cite{Calogero3}
($\alpha_{_{ij}}=0,\,f_{_j}({\vec\rho})=\sum_{ik}\lambda_{_{jik}}\,\rho_{_i}\,\rho_{_k}$).
Moreover, for $q=p=2$, with $b_{_{ij}}+2\,d_{_{ij}}=0$ and
$f_{_1}({\vec\rho})=f\,\rho_{_1}+g\,\rho_{_2}$,
$f_{_2}({\vec\rho})=g\,\rho_{_1}+f\,\rho_{_2}$, equation
(\ref{example}) has been studied in \cite{Tsuchida}.\\
The canonical sub-family of equation (\ref{example}) is given by
$b_{_{ij}}=c_{_{ji}}=-2\,d_{_{ij}}=-2\,e_{_{ij}}$ and can be
obtained through the following nonlinear potential
\begin{equation}
U[{\vec\rho},\,{\vec
S}]=-\sum_{i,j=1}^pb_{_{ij}}\,\rho_{_i}\,\rho_{_j}\frac{\partial
S_{_i}}{\partial x} +F({\vec\rho}) \ ,
\end{equation}
where the conditions $\delta\,F({\vec\rho})/\delta\,\rho_{_j}=f_{_j}({\vec\rho})$ are assumed.\\
We observe that:\\
a) when $d_{_{ij}}=e_{_{ij}}$, for $i\not= j$, equation
(\ref{example}) conserves the densities $\rho_{_j}$ and the currents
take the form
\begin{equation}
j_{_{\psi,j}}=2\,a_{_j}\,\rho_{_j}\frac{\partial S_{_j}}{\partial
x}-(d_{_{jj}}+e_{_{jj}})\,\rho_{_j}^2-2\,
\sum_{i=1,i\not=j}^pd_{_{ij}}\,\rho_{_i}\,\rho_{_j} \ ,\label{jj}
\end{equation}
with ${\mathcal
J}_{_j}({\vec\rho})=-(d_{_{jj}}+e_{_{jj}})\,\rho_{_j}^2-2\,\sum_{i\not=
j}
d_{_{ij}}\,\rho_{_i}\,\rho_{_j}$ and $I_{_j}({\vec\rho})=0$.\\
b) when $d_{_{ij}}+e_{_{ji}}=d_{_{ji}}+e_{_{ij}}$, equation
(\ref{example}) conserves the total density $\rho_{_{\rm
tot}}=\sum_j\rho_{_j}$, and the total current is given by
\begin{equation}
j_{_{\rm tot}}=\sum_{j=1}^p\left[2\,a_{_j}\,\rho_{_j}\frac{\partial
S_{_j}}{\partial x}
-\sum_{i=1}^p(d_{_{ij}}+e_{_{ji}})\,\rho_{_i}\,\rho_{_j}\right] \ ,
\end{equation}
with ${\mathcal
J}_{_j}[{\vec\rho}]=-(d_{_{jj}}+e_{_{jj}})\,\rho_{_j}^2$ and
\begin{equation}
I_{_{j}}[{\vec\rho}]=-2\sum_{i\not=j}\left(d_{_{ij}}\,
\rho_{_i}\frac{\partial\rho_{_j}}{\partial x}+e_{_{ij}}\,\rho_{_j}\,
\frac{\partial\rho_{_i}}{\partial x}\right) \ .
\end{equation}
If we choose the functionals ${\mathcal R}_{_j}[{\vec\rho}]=0$ in
the case a) and
\begin{equation}
{\mathcal R}_{_j}[{\vec\rho}]=-\sum_{i=1,\,i\not=j}^p
\lambda_{_{ij}}\,\rho_{_i}\,\rho_{_j} \ ,
\end{equation}
in the case b), where $\lambda_{_{ij}}=d_{_{ij}}+e_{_{ij}}$, the
generators (\ref{cgenf}) can be written in the unified form
\begin{equation}
\sigma_{_j}[{\vec\rho}]=-{1\over
2\,a_{_j}}\sum_{i=1}^p\lambda_{_{ij}}\int\limits^x\rho_{_i}\,dx^\prime
\ .
\end{equation}
By performing the gauge transformation, from equation
(\ref{example}) we obtain a new system of CNLSEs for the field
$\Phi$ with nonlinearity
\begin{equation}
\widehat W^\prime[{\vec\rho},\,{\vec S}]=\widehat
D[{\vec\rho},\,{\vec S}]+\widehat C[{\vec\rho},\,{\vec S}] \ ,
\end{equation}
where the diagonal matrix $\widehat D[{\vec\rho},\,{\vec S}]$ has
entries
\begin{eqnarray}
\nonumber \widehat D[{\vec\rho},\,{\vec S}]={\rm
diag}&&\left[\sum_{i=1}^p\rho_{_i}\,\left(\mu_{_{ij}}\,\frac{\partial{\mathcal
S}_{_j}}{\partial x}+\nu_{_{ij}}\, \frac{\partial{\mathcal
S}_{_i}}{\partial
x}\right)\right.\\
&&\left.+\sum_{i,k=1}^p\omega_{_{jik}}\,\rho_{_i}\,\rho_{_k}+f_{_j}({\vec
\rho})\right] \ ,
\end{eqnarray}
with
\begin{eqnarray}
\nonumber
&&\mu_{_{ij}}=b_{_{ij}}+\lambda_{_{ij}} \ ,\\
&&\nu_{_{ij}}=c_{_{ij}}-\frac{a_{_i}}{a_{_j}}\,\lambda_{_{ij}} \ ,\\
\nonumber
&&\omega_{_{jik}}={1\over4\,a_{_j}}\left(\lambda_{_{ij}}\,\lambda_{_{kj}}+2\,b_{_{ij}}\,
\lambda_{_{kj}}+2\,\frac{a_{_j}}{a_{_i}}\,c_{_{ij}}\,\lambda_{_{ki}}\right)
\ ,
\end{eqnarray}
whereas the off-diagonal matrix $\widehat C$ has entries
\begin{equation}
\left(\widehat C\right)_{_{ij}}\!\!\![{\vec\rho},\,{\vec
S}]=i\,\frac{{\mathcal F}_{_i}({\vec\rho})-{\mathcal
F}_{_j}({\vec\rho})}{2\,p\,\sqrt{\rho_{_i}\,\rho_{_j}}}\,e^{i\,\left({\mathcal
S}_{_i}-{\mathcal S}_{_j}\right)} \ ,
\end{equation}
where
\begin{equation}
{\mathcal F}_{_j }({\vec\rho})=\sum_{i=1}^p(d_{_{ij}}-e_{_{ij}})\,
\left(\rho_{_i}\,\frac{\partial\rho_{_j}}{\partial x}
-\frac{\partial\rho_{_i}}{\partial x}\,\rho_{_j}\right) \
.\label{ff}
\end{equation}
We observe that the functionals (\ref{ff}) vanish in the case a) and
the nonlinearity
$\widehat W^\prime[{\vec\rho},\,{\vec S}]$ reduces to a purely real quantity.\\
Let us now collect some particular cases belonging to equation
(\ref{example}).\\
1) By choosing $b_{_{ij}}=-\lambda_{_{ij}}$ and
$a_{_j}\,c_{_{ij}}=2\,a_{_i}\,\lambda_{_{ij}}$ we obtain a system of
CNLSEs with a purely real nonlinearity which depends only on the
fields $\rho_{_i}$
\begin{equation}
i\,\frac{\partial\phi_{_j}}{\partial
t}+a_{_j}\,\Delta\phi_{_j}-\left(\sum_{i,k=1}^p\omega_{_{jik}}\,
\rho_{_i}\,\rho_{_k}+f_{_j}({\vec\rho})\right)\,\psi_{_j}=0 \
.\label{ris1}
\end{equation}
When
$f_{_j}(\vec\rho)=\sum_{ik}\lambda_{_{jik}}\,\rho_{_i}\,\rho_{_k}$
 with
$\lambda_{_{jik}}=\sum_{_{ik}}b_{_{ij}}(b_{_{kj}}-2\,b_{_{ki}})/4\,a_{_j}$,
it reduces to a system of decoupled linear Schr\"odinger equations
\begin{equation}
i\,\frac{\partial\phi_{_j}}{\partial t}+a_{_j}\,\Delta\phi_{_j}=0 \
.
\end{equation}
2) By choosing $b_{_{ij}}=-\lambda_{_{ij}}$ for $i\not=j$,
$a_{_j}\,c_{_{ij}}=2\,a_{_i}\,\lambda_{_{ij}}$ and
$f_{_j}(\vec\rho)=\sum_{ik}\lambda_{_{jik}}\,\rho_{_i}\,\rho_{_k}$
with
\begin{eqnarray}
\left\{
\begin{array}{l}
\lambda_{_{kkk}}=\lambda_{_{kk}}\,\left(b_{_{kk}}+3\,
\lambda_{_{kk}/2}\right)/2\,a_{_k}
\ ,\\
\lambda_{_{kjk}}=\lambda_{_{kj}}\left(b_{_{kk}}+
\lambda_{_{kk}}/2+\lambda_{_{jk}}\right)/2\,a_{_k}
\ ,\\
\lambda_{_{kki}}=\lambda_{_{kk}}\,\lambda_{_{ki}}/4\,a_{_k} \ ,\\
\lambda_{_{kji}}=\lambda_{_{kj}}\left(\lambda_{_{ji}}-
\lambda_{_{ki}}/2\right)/2\,a_{_k}, \quad\mbox{ for $k\not=j\not=i$ and
$\kappa\not=j=i$} \ ,
\end{array}
\right.
\end{eqnarray}
we obtain the following system of decoupled Jackiw-like NLSEs
\begin{equation}
i\,\frac{\partial\phi_{_j}}{\partial t}+a_{_j}\,
\frac{\partial^2\phi_{_j}}{\partial x^2}
+\eta_{_j}\,j_{_j}\,\phi_{_j}=0 \ ,\label{ex1}
\end{equation}
with $\eta_{_j}=(b_{_{jj}}+\lambda_{_{jj}})/2\,a_{_j}$.\\
3) By choosing $b_{_{ij}}=-\lambda_{_{ij}},\,\,\lambda_{_{kji}}=
c_{_{kj}}\,\lambda_{_{ji}}/2\,a_{_j}-\lambda_{_{kj}}\,\lambda_{_{ki}}/4\,a_{_k}$
we obtain the CNLSEs
\begin{equation}
i\,\frac{\partial\phi_{_j}}{\partial t}+a_{_j}\,
\frac{\partial^2\phi_{_j}}{\partial x^2}
+\sum_k\eta_{_{jk}}\,j_{_k}\,\phi_{_j}=0 \ ,\label{34}
\end{equation}
being $\eta_{_{jk}}=(c_{_{jk}}-a_{_k}\,
\lambda_{_{jk}}/a_{_j})/2\,a_{_k}$. The nonlinear term in equation
(\ref{34}) has been considered in \cite{Calogero5}.


\subsection{Gauged NLSEs}

Let us consider a system of charged particles undergoing to
anomalous diffusion and described by the following NLSE
\begin{equation}
i\,D_t\psi+{\bfm D}^2\psi+\Lambda[\rho,\,S,\,{\bfm A}]\,\psi=0 \
,\label{exe6}
\end{equation}
with nonlinearity
\begin{eqnarray}
\nonumber \Lambda[\rho,\,S,\,{\bfm
A}]&&=\left[a_{_1}\,\frac{{\bfm\nabla}\cdot\left({\bfm\nabla}S-{\bfm
A}\right)}{\rho^{1-q}}+a_{_2}\,\frac{\Delta\rho}{\rho^{3-2\,q}}+
a_{_3}\,\left(\frac{{\bfm\nabla}\rho}{\rho^{2-q}}\right)^2\right]\\
&&+i\,\frac{D}{2}\,\frac{\Delta\rho^q}{\rho} \ , \label{lecm}
\end{eqnarray}
where $a_{_1}=q\,D$, $a_{_2}=2\,\alpha$ and
$a_{_3}=\alpha\,(2\,q-3)$ with $\alpha,\,q$ and $D$ constant
parameters. Equation (\ref{exe6}) must be considered jointly with
\begin{equation}
\gamma\,\partial_\mu F^{\mu\nu}+g\,\varepsilon^{\nu\tau\mu}
\,F_{\tau\mu}=j^\nu_{_{A\psi}} \ ,\label{exe61}
\end{equation}
describing the dynamics of the gauge field.\\
The nonlinearity (\ref{lecm}) can be obtained from the potential
\begin{equation}
U[\rho,\,S,\,{\bfm A}] =D\,q
\,\rho^{q-1}\,{\bfm\nabla}\rho\cdot({\bfm\nabla}S-{\bfm
A})+\alpha\,\rho^{2q-3}\,\frac{({\bfm\nabla}\rho)^2}{\rho} \
,\label{pot}
\end{equation}
and the charged current ${\bfm j}_{_{A\psi}}$ is given by
\begin{equation}
{\bfm j}_{_{A\psi}}={\bfm
j}^{(0)}_{_{A\psi}}+D\,q\,\rho^{q-1}\,{\bfm\nabla}\rho \ ,
\end{equation}
with ${\bfm j}^{(0)}_{_{A\psi}}=2\,\rho\,({\bfm\nabla}S-{\bfm A})$.\\
As a consequence the system fulfills the following continuity
equation
\begin{equation}
\frac{\partial\rho}{\partial t}+{\bfm\nabla}\cdot{\bfm
j}_{_{A\psi}}^{(0)}+D\,\Delta\rho^q=0 \ ,\label{cexe6}
\end{equation}
which is a nonlinear Fokker-Planck equation for charged particles.\\
By performing the transformation (\ref{trasf1g}) with
\begin{equation}
\sigma[\rho]=\frac{D}{2}\,\frac{q\,\rho^{q-1}-1}{q-1} \
,\label{trexe6}
\end{equation}
where the integration constant has been chosen to avoid the
singularity for $q\to1$, equations (\ref{exe6}) and (\ref{exe61})
are transformed in
\begin{equation}
i\,D_t\phi+{\bfm D}^2\phi+\beta\,\rho^{2q-2}\,\left[
\frac{\Delta\rho}{\rho}+\left(q-{3\over2}\right)
\,\left(\frac{{\bfm\nabla}\rho}{\rho}\right)^2 \right]\,\phi=0 \ ,
\label{dob2}
\end{equation}
with $\beta=2\,\alpha-q^2\,D^2/2$ and
\begin{equation}
\gamma\,\partial_\mu F^{\mu\nu}+g\,\varepsilon^{\nu\tau\mu}
\,F_{\tau\mu}=j^\nu_{_{A\phi}} \ ,\label{dob21}
\end{equation}
where $j^\nu_{_{A\phi}}=(\rho,\,{\bfm j}^{(0)}_{_{A\phi}})$ with
${\bfm j}^{(0)}_{_{A\phi}}=2\,\rho\,({\bfm\nabla}{\mathcal S}-{\bfm
A})$.\\
Similar equations can be obtained equivalently by means of the
transformation
\begin{eqnarray}
{\bfm\chi}&=&{\bfm
A}-\frac{D\,q}{2}\,\rho^{q-2}\,{\bfm\nabla}\rho \ ,\\
\chi_0&=&A_0-\frac{D\,q}{2}\,\rho^{q-2}\,{\bfm\nabla}\cdot{\bfm
j}_{_{A\psi}} \ .\label{trgauge}
\end{eqnarray}
It is worthy to observe that equation (\ref{exe6}), in the $q\to1$
limit, reduces to the gauged canonical subclass of the
Doebner-Goldin family discussed in section 6.1
\begin{eqnarray}
\nonumber i\,D_t\psi+{\bfm D}^2\psi
&+&\left[D\,{\bfm\nabla}\cdot\left({\bfm \nabla}S-{\bfm
A}\right)+2\,\alpha\,\frac{\Delta\rho}{\rho}-\alpha\,
\left(\frac{{\bfm\nabla}\rho}{\rho}\right)^2\right]\,\psi\\
&+&i\,{D\over2}\,\frac{\Delta\rho}{\rho}\,\psi=0 \ ,\label{exe62}
\end{eqnarray}
which is obtainable from the potential
\begin{equation}
U[\rho,\,S,\,{\bfm A}]
=D\,{\bfm\nabla}\rho\cdot({\bfm\nabla}\,S-{\bfm A})+
\alpha\,\frac{({\bfm\nabla}\rho)^2}{\rho} \ ,
\end{equation}
and the continuity equation (\ref{cexe6}) reduces to the linear
Fokker-Planck equation for charged particles
\begin{equation}
\frac{\partial\rho}{\partial
t}+{\bfm\nabla}\cdot\bfm{j}_{_{A\psi}}^{(0)}+D\,\Delta\rho=0 \
.\label{ccexe61}
\end{equation}
In the same limit the gauge transformation has generator
\begin{equation}
\sigma[\rho]={D\over2}\,\log\rho \ ,
\end{equation}
and reduces equation (\ref{exe62}) to
\begin{equation}
i\,D_t\phi+{\bfm D}^2\phi+\beta \left[\frac{\Delta\rho}{\rho}
-{1\over2}\,\left(\frac{{\bfm\nabla}\rho}{\rho}\right)^2\right]\,\phi=0
\ ,\label{dob61}
\end{equation}
with $\beta=2\,\alpha-D^2/2$.

\sect{Conclusions and comments}

In this paper we have considered a class of canonical NLSEs
containing complex nonlinearities and describing U(1)-invariant
systems. We have introduced a unitary and nonlinear transformation
$\psi\rightarrow\phi$ which reduces the complex nonlinearity in a
real one and at the same time transforms the quantum particles
current in the standard bilinear form. We have extended the method
to U(1)-invariant CNLSEs. For these systems we have generalized the
gauge transformation with the purpose to change the initial
nonlinearity in another one purely Hermitian. It has been shown that
there are many different possibilities to define the generator of
the transformation. For any choice we obtain a new CNLSE with a
different, but Hermitian, nonlinearity.  Finally, we have
specialized the method for NLSEs minimally coupled with an Abelian
gauged field. We have shown that there are two different ways to
reduce the complex nonlinearity in a purely real one: or by a
nonlinear unitary transformation on the matter field or,
alternatively, by a nonlinear transformation on the gauge field.\\
In the following let us make some considerations about the
transformation studied in the present work.\\
Firstly, the problem of the integrability of a nonlinear evolution
equation is one of the most studied topics in mathematical physics.
Let us consider the most general U(1)-invariant scalar NLSE in the
hydrodynamic representation
\begin{eqnarray}
&&\frac{\partial\rho}{\partial
t}+{\bfm\nabla}\cdot\left(2\,\rho\,{\bfm
\nabla}\,S+{\bfm{\mathcal J}}\right)=0 \ ,\label{hjcb}\\
&&\frac{\partial S}{\partial t}+ ({\bfm\nabla}S)^2+U_{_q}-W=0 \
.\label{hjca}
\end{eqnarray}
In the Calogero picture \cite{Calogero3,Calogero4a,Calogero4,Calogero1}, the
system of equations (\ref{hjcb}) and (\ref{hjca}) is $C$-integrable
if there exists a transformation of the dependent and/or independent
variables: $t\rightarrow T,\,\,{\bfm x}\rightarrow{\bfm
X},\,\,\rho\rightarrow R,\,\,S\rightarrow{\mathcal S}$ which changes
equations (\ref{hjcb}), (\ref{hjca}) in
\begin{eqnarray}
&&\frac{\partial R}{\partial T}+
\overline{\bfm\nabla}\cdot\left(2\,R\,\overline{\bfm\nabla}{\mathcal
S}\right)=0 \ ,\label{hjc1b}\\ &&\frac{\partial{\mathcal
S}}{\partial T}+(\overline{\bfm\nabla}{\mathcal S})^2
+\overline{U}_q=0 \ ,\label{hjc1a}
\end{eqnarray}
where $\overline{\bfm\nabla}$ and $\overline{U}_q$ are the gradient
and the quantum potential in the new variables. Equations
(\ref{hjc1b}) and (\ref{hjc1a}) constitute the well known
hydrodynamic representation of the standard linear Schr\"odinger
equation.\\
On the other hand, the transformation on the field
$S\rightarrow{\mathcal S}$ introduced in this paper, reduces the
continuity equation (\ref{hjcb}) in the standard form given by
equation (\ref{hjc1b}) and can be seen as a first step in the
Calogero program.\\
Secondly, the most general gauge transformation of the kind
discussed in the present work can be stated as
\begin{equation}
\psi(t,\,{\bfm x})\to\phi(t,\,{\bfm x})={\mathcal
U}[\rho,\,S]\,\psi(t,\,{\bfm x}) \ ,\label{trtrr}
\end{equation}
which is an infinite dimensional unitary representation of the
diffeomorphism group with
\begin{equation}
{\mathcal U}[\rho,\,S]=\exp\Big(i\,\omega[\rho,\,S]\Big) \
.\label{trtr}
\end{equation}
As matter of fact the real generator $\omega[\rho,\,S]$ could be any
arbitrary functional depending
on the fields $\rho$ and $S$.\\
For instance, in \cite{Doebner4} the generator of the transformation
has been assumed in
\begin{equation}
\omega(\rho,\,S)=\frac{\gamma(t)}{2}\,\log{\rho}+
(\lambda(t)-1))\,S+\theta(t,\,{\bfm x}) \ ,\label{dbt}
\end{equation}
which produces a group of transformations mapping the Doebner-Goldin
equation in itself. We observe that the one parameter subclass of
this transformation with $\theta(t,{\bfm
x})=0,\,\lambda(t)=1$ and $\gamma(t)=constant$, coincides with those studied in this work.\\
Throughout this paper, the generator of the gauge transformation has
been chose with the purpose to make real the complex nonlinearity of
the NLSE under inspection. Alternatively, nonlinear gauge
transformations can be useful generalized with the purpose to
classify NLSEs in equivalence classes. Any equation belonging to the
same class, in spite of its nonlinearity, is gauge equivalent, by
means of equation (\ref{trtr}), to the others equations belonging to
the same class. \\
For instance, let us consider the following family of NLSEs
\begin{equation}
i\,\frac{\partial\psi}{\partial
t}+\Delta\psi+\Lambda[\rho,\,S]\,\psi=0 \ ,\label{eqeq}
\end{equation}
with complex nonlinearity
\begin{eqnarray}
\nonumber \Lambda[\rho,\,S]&=&f_{_1}(\rho)\,\Delta
S+f_{_2}(\rho)\,{\bfm\nabla}\rho\cdot{\bfm\nabla}S+f_{_3}(\rho)\,({\bfm\nabla}\rho)^2
\\&+&f_{_4}(\rho)\,\Delta\rho+{i\over\rho}\,{\bfm\nabla}\Big(f_{_5}(\rho)\,{\bfm\nabla}\rho\Big)
\ ,\label{cnn}
\end{eqnarray}
where the expression of the imaginary part guarantees the existence
of a continuity equation for $\rho$.\\ The quantities $f_{_i}(\rho)$
are functional parameters fixing the NLSE. Any NLSE belonging to the
family of equations (\ref{eqeq}) can be determined univocally
through the vector $\vec
f\equiv\{f_{_1}(\rho),\,\ldots,\,f_{_5}(\rho)\}$. By performing a
gauge transformation with generator $\omega(\rho)$ depending only on
the field $\rho$, equation (\ref{eqeq}) changes in
\begin{equation}
i\,\frac{\partial\phi}{\partial
t}+\Delta\phi+\widetilde\Lambda[\rho,\,{\mathcal S}]\,\phi=0 \
,\label{cnnnn}
\end{equation}
with
\begin{eqnarray}
\widetilde\Lambda[\rho,\,S]&=&\tilde f_{_1}(\rho)\,\Delta {\mathcal
S}+\tilde f_{_2}(\rho)\,{\bfm\nabla}\rho\cdot{\bfm\nabla}{\mathcal
S}+\tilde f_{_3}(\rho)\,({\bfm\nabla}\rho)^2\\ &+&\tilde
f_{_4}(\rho)\,\Delta\rho+{i\over\rho}\,{\bfm\nabla}\Big(\tilde
f_{_5}(\rho)\,{\bfm\nabla}\rho\Big) \ .\label{cn}
\end{eqnarray}
What is important is to note that the transformation maintains the
same structure in the nonlinearity through the presence of the
functional groups $\Delta
S,\,{\bfm\nabla}S\cdot{\bfm\nabla}\rho,\,({\bfm\nabla}\rho)^2$ and
$\Delta\rho$, whilst the expressions of the new parameters $\tilde
f_{_i}(\rho)$ are given by
\begin{eqnarray}
\nonumber
&&\widetilde f_{_1}=f_{_1}-2\,\rho\,\frac{\partial\omega}{\partial\rho} \ ,\\
\nonumber
&&\widetilde f_{_2}=f_{_2} \ ,\\ \nonumber&&\widetilde
f_{_3}=f_{_3}-\left[f_{_2}-\frac{\partial\omega}{\partial\rho}
+2\,\frac{\partial
f_{_5}}{\partial\rho}+\left(f_{_1}-2\,\rho\,\frac{\partial\omega}{\partial\rho}\right)
\,\frac{\partial}{\partial\rho}\right]\frac{\partial\omega}{\partial\rho}
 \ ,\\
 \nonumber
&&\widetilde f_{_4}=f_{_4}-\left(f_{_1}
+2\,f_{_5}-2\,\rho\,\frac{\partial\omega}{\partial\rho}\right)\frac{\partial\omega}{\partial\rho}
\ ,\\
&&\widetilde f_{_5}=f_{_5}-\rho\,\frac{\partial\omega}{\partial\rho}
\ .
\end{eqnarray}
By eliminating $\omega(\rho)$ among these equations we obtain a set
of gauge invariants relations
\begin{eqnarray}
\nonumber
&&\widetilde f_{_1}-f_{_1}=2\,\Big(\widetilde
f_{_5}-f_{_5}\Big) \ ,\label{1}\\
\nonumber
&&\widetilde f_{_3}-f_{_3}={1\over\rho}\left[f_{_2}+
{1\over\rho}\,\Big(\widetilde
f_{_5}-f_{_5}-f_{_1}\Big)+2\,\frac{\partial\widetilde
f_{_5}}{\partial\rho}
+f_{_1}\,\frac{\partial}{\partial\rho}\right]\Big(\widetilde
f_{_5}-f_{_5}\Big) \ ,\\
&&\widetilde f_{_4}-f_{_4}={1\over\rho}\,\Big(f_{1}+2\,\widetilde
f_{_5}\Big)\,\Big(\widetilde f_{_5}-f_{_5}\Big) \ .\label{3}
\end{eqnarray}
Given two NLSEs belonging to the family (\ref{eqeq}), labeled by the
respective vectors $\vec
f\equiv\{f_{_1}(\rho),\,\ldots,\,f_{_5}(\rho)\}$ and $\vec
f^\prime\equiv\{\tilde f_{_1}(\rho),\,\ldots,\,\tilde
f_{_5}(\rho)\}$, if the functionals $f_{_i}(\rho)$ and $\tilde
f_{_i}(\rho)$ fulfil the relations (\ref{3}), the two
NLSEs are gauge equivalents since there exist a generator
$\omega(\rho)$ such that, by means of equation (\ref{trtrr}),
transforms the first NLSE,
labeled by the vector $\vec f$, in the second NLSE labeled by the vector $\vec f^\prime$.\\
In particular, observing that the linear Schr\"odinger equation is
represented by the vector $\vec f\equiv\{0,\,0,\,0,\,0,\,0\}$, it
follows that any NLSE fulfilling the relations
\begin{eqnarray}
\nonumber
&&f_{_1}=2\,f_{_5} \ ,\label{11}\\
\nonumber
&&f_{_2}=0 \ ,\\
\nonumber
&&f_{_3}={f_{_5}\over\rho}\left(2\,\frac{\partial
f_{_5}}{\partial\rho}-{f_{_5}\over\rho}\right) \
,\\
&&f_{_4}=2\,{f_{_5}^2\over\rho} \ ,\label{14}
\end{eqnarray}
is gauge equivalent to the linear Schr\"odinger equation. This sub-family can be
linearizable by means of the transformation (\ref{trtrr}) with
generator
$\omega(\rho)=\int^\rho(f_{_5}(\rho^\prime)/\rho^\prime)\,d\rho^\prime$.
In this sense, equations (\ref{14}) define the subclass
of the family of equations (\ref{eqeq}) which are $C$-integrable.\\
In conclusion, we have shown  that the transformation introduced in
the present work allows us to deal, in a unifying scheme, different
NLSEs already known in literature, obtaining in a
systematic way the transformations introduced by various Authors. \\
A natural continuation of this work could be performed in
several ways:\\
1) Extending the method to the case of NLSEs coupled with non
Abelian gauge fields, which are relevant, for instance, in the study of heavy-quark particle systems.\\
2) Extending the method to relativistic nonlinear equations. In this
context, in \cite{Doebner5}, a relativistic generalization of the
transformation introduced in \cite{Doebner4} has been proposed to
generate
nonlinear extensions of the Dirac equation.\\
3) Extending the method to discrete NLSEs, which are particularly
relevant in the study of lattice models in condensed matter.

\vfill\eject

\end{document}